\documentclass[%
 reprint,
 amsmath,amssymb,
 aps,
nofootinbib]{revtex4-2}

\usepackage{graphicx}
\usepackage{dcolumn}
\usepackage{bm}
\usepackage{newtxtext,newtxmath}
\usepackage{lipsum}
\usepackage{xcolor}
\usepackage{mathtools} 
\usepackage{lipsum}
\usepackage{natbib}

\usepackage[T1]{fontenc}

\usepackage{hyperref}
\hypersetup{
    colorlinks=true,
    linkcolor=blue,
    citecolor = blue,
    filecolor=blue,      
    urlcolor=magenta,
    pdftitle={Torsional oscillations of magnetized neutron stars},
    pdfpagemode=FullScreen,
    }

\begin{document}

\preprint{APS/123-QED}

\title{Torsional oscillations of magnetized neutron stars: \\Impacts of Landau-Rabi quantization of electron motion}

\author{Ling Cheung}
\email{lingcheungug@link.cuhk.edu.hk}
\author{Lap-Ming Lin}%
\affiliation{%
 Department of Physics, The Chinese University of Hong Kong, Hong Kong, China
}%


\author{Nicolas Chamel}
\affiliation{
 Institute of Astronomy and Astrophysics,
Universit\'{e} Libre de Bruxelles, CP 226,
Boulevard du Triomphe, B-1050 Brussels, Belgium
}%

\date{\today}

\begin{abstract}
Torsional oscillations of magnetized neutron stars have been well studied since they may be 
relevant to the physical interpretation of some of the observed quasiperiodic oscillations in the magnetar 
giant flares. In the crustal region of a magnetar, the strong magnetic field can alter the equation 
of state and composition due to the Landau-Rabi quantization of electron motion. 
In this paper, we study this effect on the crust-confined, torsional oscillation modes of neutron stars with 
mixed poloidal-toroidal magnetic fields in general relativity under the Cowling approximation.
Furthermore, the inner and outer crusts are treated consistently based on the nuclear-energy density 
functional theory. 
Depending on the magnetic-field configurations, we find that the Landau-Rabi quantization of electrons can change the frequencies of the fundamental torsional oscillation mode of $1.4 M_\odot$ neutron star models with a normal fluid core by about 10\% when the magnetic field strength at the pole reaches the order of $10^{16}$ G. 
The shift can even approach 20\% at a field strength of $10^{15}$ G for neutron stars with a simple model of superconducting core where the magnetic field is assumed to be expelled completely. 
\end{abstract}

\maketitle

\section{Introduction}
\label{sec:intro}

Neutron stars, which are born from the gravitational core-collapse of massive stars during supernova explosions, contain a superdense core that can reach several times the nuclear matter density \citep{ShapiroTeukolsky1983}. 
It is now well established that highly magnetized neutron stars, magnetars, with magnetic field strengths reaching over $10^{14} - 10^{15}$ G can exist, making them the most magnetic objects in the universe \citep{Turolla_2015}.  
They are responsible for the observed soft-gamma ray repeaters (SGRs) and 
anomalous X-ray pulsars \citep{Kaspi:2017}. 
High energy sudden X-ray outbursts with luminosities up to $\sim 10^{36} {\rm erg \ s}^{-1}$ have 
been observed for some of these objects \citep{Rea:2011}. More energetic giant flares with peak 
luminosities $10^{44} - 10^{47} {\rm erg \ s}^{-1}$ have also been observed 
\citep{Hurley:1999,Hurley:2005,Mereghetti:05}. 

Quasiperiodic oscillations (QPOs) with a large range of frequencies have been detected 
in the late-time tail phase of giant flares. Examples for such frequencies are 18, 26, 29, 92.5, 150, 626.5 and 1837 Hz for SGR 1806-20 \citep{Israel1806_20,Watts_2006,Watts1806_20}; 28, 54, 84 and 155 Hz for SGR 1900+14 \citep{Watts1900+14} and 43.5 Hz for SGR 0526-66 \citep{Barat0526_66}. More recently, very high-frequency QPOs at 2132 and 4250 Hz have also been identified in the main peak of a giant flare \citep{Castro:2021}. 

While the origin of the QPOs is still an open question, they are commonly associated to the magneto-elastic oscillations of the star. 
The fundamental theory for torsional oscillations of elastic solid crusts in the absence of magnetic fields was developed in \citep{Hansen1980,SchumakerThorne1983,McDermott1988}.
Models focusing on the torsional oscillations of the solid crust with dipole magnetic fields were investigated in \citep{Sotani2007}, or even with mixed poloidal-toroidal fields in \citep{de_Souza_2019,Sotani_2008_phi}. 
Torsional oscillations might explain some but not all observed QPOs. Complications due to the coupling of the crust to the Alfv\'{e}n modes in the fluid core are still an unsettled issue \citep{Levin:2006,Gabler:2011}. 
Nevertheless, oscillations of highly magnetized neutron stars have been extensively investigated in numerous publications. Some studies have taken into account the intricate coupling to the magnetic field and have discussed
the qualitative changes of the QPOs from shear-like to Alfv\'{e}n-like behavior (e.g., \citep{Levin:2006,Gabler:2011,Levin2007,Glampedakis2006,Sotani2007,2009MNRAS.397.1607C,Colaiuda2009,Colaiuda2011,Colaiuda2012,vanHoven2011,vanHoven2012}). 
Among these studies, the global crust-core oscillations of highly magnetized neutron stars 
was discussed by \citep{Levin:2006,Glampedakis2006}. 
The continuum nature of QPOs spectrum was pointed out in \citep{Levin2007} using a toy model (see also \citep{Sotani2008_AlfvenQPO,Colaiuda2009,Colaiuda2011} for further discussions). 
However, it was later found that the axial spectrum loses its continuum character when the coupling between the polar and axial oscillations is taken into account \citep{Colaiuda2012}. 
Moreover, it was demonstrated that the presence of the neutron star crust has a significant impact on the continuum spectrum. Specifically, it gives rise to the existence of discrete Alfv\'{e}n modes near the edges of the continua, which correspond to the turning points of the continuum \citep{2010JPhCS.222a2031K,vanHoven2011,Gabler:2011,Gabler:2012}.  
Despite extensive efforts, the explanation for the high-frequency QPOs at approximately 625 Hz has proven challenging for neutron stars with nonsuperfluid core \citep{vanHoven2012}. 
This is due to the fact that at high frequencies, the continua dominate and overlap with each other, effectively absorbing the discrete frequencies. Consequently, it becomes very difficult to identify the edges of the continuum, further complicating the interpretation of the observed QPOs.
Nevertheless, it has been suggested that the high-frequency QPOs can be interpreted 
in terms of global magneto-elastic oscillations when the effects of superfluidity are
taken into account \citep{2013PhRvL.111u1102G,2014MNRAS.438..156P,2018MNRAS.476.4199G}.
An attempt was also made to investigate the polar oscillations of magnetars (without crust) in an effort to 
understand the high-frequency QPOs \citep{Sotani2009_alfvenpolar} as the polar Alfv\'{e}n oscillations was found to be discrete, in contrast to the axial Alfv\'{e}n oscillations. The frequencies of the fundamental polar Alfv\'{e}n oscillations were found to be on the order of a few hundred Hz for typical magnetars.
On the other hand, the low-frequency QPOs may be relevant to torsional Alv\'{e}n modes \citep{Sotani2008_AlfvenQPO}.
The study of the magneto-elastic oscillations of magnetars touches upon various areas of physics, ranging
from the underlying magnetic field configuration to the physics of dense matter, and so far there is
no self-consistent model that can incorporate all the important physics components in the analysis. 

The intense magnetic field also affects the properties of the crust. 
In particular, one important effect is due to the quantization of the electron motion perpendicular 
to the field into Landau-Rabi levels \citep{rabi1928freie,landau1930diamagnetismus}. 
As a consequence, the composition and equation of state (EOS) of the crustal region can be modified significantly in a high magnetic field \citep{Chamel_2012,PhysRevC.91.065801,Mutafchieva_2019}.   
The role of Landau-Rabi quantization of electron motion has recently been studied by treating the inner and outer crusts consistently based on the nuclear-energy density functional theory \citep{Mutafchieva_2019}.
More specifically, the outer crust is described using the experimental atomic mass data \cite{kondev2017ame2016} supplemented
with the Brussels-Montreal atomic mass model HFB-24 \citep{goriely2013further}, while the inner crust 
is calculated based on the same functional BSk24 that underlies the HFB-24 model. The EOS of the
inner and outer crusts are thus treated in a 
unified way. 
As a result, the composition and EOS of the magnetar crust are found to vary with the magnetic field due to the Landau-Rabi quantization of electron motion. 
The EOS of the crust matches smoothly that calculated in \citep{2018MNRAS.481.2994P} for the core using the same functional. It is worth noting that the application of a thermodynamically consistent and unified EOS throughout the whole star is
important as it has been recently shown that an ad hoc matching of different EOSs for the crust and core 
can lead to significant errors on the neutron-star structure \citep{Ferreira_2020,PhysRevC.104.015801}. 
The stellar radius can differ by a few percent while the thickness of the crust can differ by up to 30\%.
This can have significant effects on the crustal torsional oscillation modes considered in this work.

Previous studies (e.g., \cite{Sotani2007,Sotani:2018,Sotani:2022}) have shown that torsional oscillation mode frequencies depend sensitively on the EOS and properties of the crust, and hence one may try to constrain the EOS with observational data if these modes contribute to the QPOs observed in the giant flares of magnetars \cite{Sotani:2023}. 
In this work, we extend the study of torsional oscillation modes of magnetars by incorporating the effects of Landau-Rabi quantization of electron motion. We shall in particular focus on the errors
that will be made if one ignores the effects of magnetic field on the EOS in the oscillation mode calculation. 
Moreover, it has been proven that both purely dipole and purely toroidal magnetic field configurations are generally unstable for non-convective star (e.g., \cite{Flowers:1977,Frieben:2012}). Therefore, we shall consider mixed poloidal-toroidal magnetic fields in this work.

The paper is structured as follows. The equations governing a hydrostatic equilibrium star
in general relativity with mixed poloidal-toroidal magnetic fields are summarized in Sec.~\ref{sec:overall STATIC STELLAR BACKGROUNDS}.
In Sec.~\ref{sec:EQUATIONS OF STATE}, the magnetic-field-dependent EOS model employed in this work
is described. 
Sec.~\ref{sec:SELF-CONSISTENT STELLAR MODELS} discusses the assumptions and methods that we introduce to construct our magnetized stellar models and magnetic field configurations. 
In Sec.~\ref{sec:TORSIONAL OSCILLATIONS}, the perturbation equations for calculating the torsional
oscillation modes of magnetized neutron stars are described. 
The structures of neutron stars calculated from our magnetized stellar models are then presented in Sec.~\ref{sec:STRUCTURES OF NEUTRON STARS}. 
Our main numerical results for the torsional oscillation modes are
presented in Sec.~\ref{sec:MODE FREQUENCIES}. The effects of different stellar masses on the 
mode frequencies are discussed in Sec.~\ref{sec:Another model for comparison}. Finally, we 
summarize and discuss our results in Sec.~\ref{sec:Discuss}. Unless stated otherwise, we assume 
geometric units with $G=c=1$ and the metric signature is $(-,+,+,+)$.

\section{STATIC STELLAR BACKGROUNDS}
\label{sec:overall STATIC STELLAR BACKGROUNDS}
\subsection{Hydrostatic equilibrium}
\label{sec:STATIC STELLAR BACKGROUNDS}

In all the models presented in this work, neutrons stars are assumed to be spherical irrespective of the magnetic field  configuration. 
That is, the 
stellar deformations due to the tensions of magnetic fields are neglected. 
This treatment is expected to be a good approximation for the magnetic field strengths $B\sim 10^{15}-10^{16}$~G considered in this work since the energy of the magnetic field is considerably much smaller than the gravitational energy \citep{Sotani2007}
\begin{equation}
\frac{\textrm{magnetic\: energy}}{\textrm{gravitational\: energy}}
 \sim\frac{B^2R^3}{GM^2/R} \sim 10^{-4}\left (\frac{B}{10^{16}\,\textrm{G}}\right )^2,
\end{equation}
Detailed numerical solutions of Einstein-Maxwell equations  \cite{Chatterjee_2021} have shown that the stellar structure remains almost unchanged for magnetic fields $B\lesssim 10^{17}$~G.

In addition, magnetars are in general slowly rotating as most of the angular momentum is extracted via 
enhanced magnetic braking due to their strong magnetic fields comparing to ordinary neutron stars. 
For this reason, we ignore any rotational deformations for simplicity. The resulting equilibrium stellar models can be considered as spherically symmetric stars.
So, we consider non-rotating and non-accreting spherically symmetric neutron stars composed of cold matter. 
The spacetime for the unperturbed equilibrium stellar model is described by the static 
and spherically symmetric metric
\begin{equation}
\label{eq:metric}
ds^{2}=-e^{2\nu}dt^2+e^{2\lambda}dr^2+r^2\left ( d\theta^2+\sin^2\theta d\phi^2 \right ),
\end{equation}
where $\lambda$ and $\nu$ are two metric functions of radial coordinate $r$.
The 4-velocity of the fluid inside an unperturbed static background star is thus given by 
\begin{equation}
u^{\mu}=\left (e^{-\nu},0,0,0\right ) .
\end{equation}
 The hydrostatic equilibrium stellar model is determined by the standard Tolman-Oppenheimer-Volkoff (TOV) equation
(see, e.g., \citep{ShapiroTeukolsky1983})
\begin{equation}
\label{eq:tov p}
\frac{dP(r)}{dr}=-\frac{\left [ \rho (r)+P(r) \right ]\left [ m(r)+4\pi r^3P(r) \right ]}{r^2\left [ 1-\frac{2m(r)}{r} \right ]},
\end{equation}
\begin{equation}
\frac{dm(r)}{dr}=4\pi r^2\rho(r),
\end{equation}
where $\rho(r)$ and $P(r)$ are the energy density and pressure at coordinate radius $r$, respectively. 
The enclosed mass $m(r)$ is defined by
\begin{equation}
e^{\lambda(r)}=\frac{1}{\sqrt{1-\frac{2m(r)}{r}}} .
\end{equation}
On the other hand, the metric function $\nu(r)$ is determined by 
\begin{equation}
\frac{d \nu}{dr} = -\frac{1}{\rho} \frac{dP}{dr} \left(1 + \frac{P}{\rho} \right)^{-1} .
\end{equation}
The system is closed by providing an EOS $P(\rho)$, which will be discussed in Sec.~\ref{sec:EQUATIONS OF STATE}. The above system of equations is integrated from the center to the stellar surface with radius $R$, which is defined by $P(R)=0$. The total mass of the star is given by $M=m(R)$. Note also that the metric functions satisfy the following boundary conditions at the surface
\begin{equation}
e^{\nu(R)}=e^{-\lambda(R)}=\sqrt{1-\frac{2M}{R}}.
\end{equation}

We will describe how we construct magnetized stellar models taking into account the effects of magnetic field on the EOS in Sec.~\ref{sec:SELF-CONSISTENT STELLAR MODELS}.

\subsection{Mixed poloidal-toroidal magnetic fields}
\label{sec:MIXED POLOIDAL-TOROIDAL MAGNETIC FIELDS}

To model the mixed poloidal-toroidal magnetic fields, we employ the linearized relativistic Grad-Shafranov equation \cite{2004ApJ...600..296I}, which is derived from the Maxwell equations solved for 
static spherically symmetric stars described in Sec.~\ref{sec:STATIC STELLAR BACKGROUNDS}. 
The dipolar ($\ell=1$) component of the vector potential is related to a radial function $a_1(r)$ 
which is determined by \cite{Colaiuda_2008}  
\begin{equation}
\begin{split}
\label{eq:solve b-field}
e^{-2\lambda}\frac{d^2 a_{1} }{d r^2}+\left (\frac{d \nu}{d r}-\frac{d \lambda}{d r}\right )e^{-2\lambda}\frac{d a_{1}}{d r}+&
\left (\xi^2e^{-2\nu}-\frac{2}{r^2} \right ) a_{1}\\
&=-4\pi f_0\left (\rho+P\right )r^2 .
\end{split}
\end{equation}
where $f_0$ is an arbitrary constant and the parameter $\xi$ represents the ratio between the
toroidal and poloidal components of the magnetic field. 
The components of the magnetic field can be expressed as
\begin{equation}
\label{eq:B_component}
B_\mu = \left (0,-2\frac{e^\lambda}{r^2}a_1\cos\,\theta,e^{-\lambda}\frac{da_1}{dr}\sin\,\theta,\xi e^{-\nu}a_1 \sin^2\theta \right ).
\end{equation}
For convenience, we define $H^\mu\equiv B^\mu/\sqrt{4\pi}$ which will be used in 
Eqs.~(\ref{eq:magnetic induction equation}) and (\ref{eq:energy momentum tensor}).

Outside the star (vacuum), Eq.~(\ref{eq:solve b-field}) can be solved 
analytically. The general solution is given by \cite{Colaiuda_2008}
\begin{equation}
\label{eq:general solution outside the star}
a^{\textrm{(ex)}}_1=-\frac{3\mu_b}{8M^3}r^2\left [ \ln \left ( 1-\frac{2M}{r} \right )+\frac{2M}{r}+\frac{2M^2}{r^2} \right ],
\end{equation}
where $\mu_b$ is the magnetic dipole moment observed by an observer at infinity and the model becomes purely dipole outside the star
($\xi=0$). We require the interior solutions $a_1$ and $da_1/dr$ to match with 
Eq.~(\ref{eq:general solution outside the star}) at the stellar surface continuously. 
In the interior, $a_1$ is determined by solving Eq.~(\ref{eq:solve b-field}) numerically.

One may see from Eq.~(\ref{eq:B_component}) that there is a discontinuity in $B_\phi$ 
at the surface when $\xi > 0$. In fact, this is due to the simplified treatment used in 
the Grad-Shafranov equation\footnote{Although such a discontinuity could be caused by the presence 
of surface currents, the physical origin of these currents remains unclear.}. In reality, the drop in $B_\phi$ 
will more likely spread out in the magnetosphere so as to match the vacuum solution continuously 
(see \cite{de_Souza_2019} for a more detailed discussion).


As pointed out in \cite{Colaiuda_2008}, there may exist some points inside the star $r=\bar{r} < R$ such that $a_1(\bar{r})=0$, and hence $B_r(\bar{r})=0$; conversely such points never exist for $r > R$. 
The locations of these points will depend on the value of $\xi$. If such a point exists, the magnetic flux would be confined inside the spherical surface with radius $\bar{r}$. 
In this case, the field lines are said to be in disjoint domains. For such magnetic field configuration, a physical interpretation is still missing and we treat it as unphysical case in this work. We shall not 
study those configurations.

In this work, two different types of cores inside neutron stars will be considered: normal fluid cores and superconducting cores. As they have different properties, their corresponding magnetic field models will also be different. 

For a normal fluid core, the magnetic field extends throughout the star. This situation has been well discussed in the literature (e.g., \cite{2004ApJ...600..296I}). 
The regularity condition at the center of the star implies that $a_1$ should have the form 
\begin{equation}
a_1=\alpha_cr^2+O\left (r^4\right ),
\end{equation}
where $\alpha_c$ is a constant determined by the junction condition with $a^{\textrm{(ex)}}_1$ at the stellar surface. Here, we 
require $a_1(r)$ to have no node 
for $r < R$ for the following ranges:
\begin{equation}
\begin{split}
        &\textrm{range I} : \,0\leq \xi\leq \xi_a,\\    
        &\textrm{range II} : \,\xi_b\leq \xi\leq \xi_c . 
        \label{eq:Bfield_parameters}
\end{split}
\end{equation}
It should be noted that the values of $(\xi_a,\xi_b,\xi_c)$ all depend on the chosen EOS and the stellar model. If $\xi$ lies inside these ranges, it is defined to be physical. We shall not consider parameters outside these
ranges. 

For a superconducting core, we only consider the simplest model for which the matter inside the core behaves as 
a type I superconductor. The magnetic field is expelled completely from the core due to the Meissner effect 
and is confined in the crust. 
In such case, we 
set $a_1$ to be zero at the core-crust interface. The regular behavior near the interface implies that $a_1$ should have the form \citep{Sotani_2008_phi}
\begin{equation}
a_1=\alpha_I\left [r - \left (R - \delta r\right )\right ]+O\left (\left [r - \left (R - \delta r\right )\right ]^2\right ),
\end{equation}
where the constant $\alpha_I$ is determined by the junction condition with $a^{\textrm{(ex)}}_1$ at the stellar surface and $\delta r$ is the thickness of the crust. Here we 
require  $a_1(r)$ to have no node for $R-\delta r< r < R$ in the following range:
\begin{equation}
    0\leq \xi\leq \xi_\gamma.
    \label{eq:Bfield_xi_gamma}
\end{equation}
Similar to the case of normal fluid core, the value of $\xi_\gamma$ depends on the EOS and the stellar model. 
If $\xi$ lies inside this range, it is defined to be physical.

\section{EQUATIONS OF STATE}
\label{sec:EQUATIONS OF STATE}

In this section, we present the EOS employed in our calculations. 
Our discussion here serves as a summary of the main results reported in \citep{Mutafchieva_2019,Chamel_2022}. 
We refer the reader to these original works and references therein for a more detailed discussion. The outer crust, the inner crust, and the core of the star are discussed as follows. 
 
\subsection{Outer crust}

The EOS for arbitrary magnetic field strength is constructed using the procedure described in 
Sec.~\ref{sec:MIXED POLOIDAL-TOROIDAL MAGNETIC FIELDS} combining the data given in Tables 2 and 3 of \cite{Chamel_2022}. In short, the experimental atomic masses from the 2016 Atomic Mass Evaluation \cite{kondev2017ame2016} with the microscopic atomic mass table HFB-24 \cite{goriely2013further} from the BRUSLIB database \cite{xu2013databases} are used.
We will briefly review the microphysics inputs here \citep{Chamel_2022}. 
For the crustal region at density below the neutron-drip point and above the ionization threshold, each crustal layer is assumed to be made of fully ionized atomic nuclei $\left (A,Z\right )$ with mass number $A$ and proton number $Z$ embedded in a relativistic electron gas.

While the pressure due to nuclei is negligible, they contribute to the energy density
\begin{equation}
    \rho_N=n_NM'(A,Z)c^2,
\end{equation}
where $n_N$ is the number density of nuclei and
$M'(A,Z)$ is the mass of nuclei including the rest mass of $Z$ electrons. In principle, $M'(A,Z)$ might also depend on the magnetic field. However, the correction is very small for the magnetic fields considered here and will thus be neglected~\cite{arteaga2011}. In the following, the magnetic field will be expressed using the dimensionless ratio $B_* \equiv B/B_{\textrm{rel}}$ with
\begin{equation}
    B_{\textrm{rel}}=\frac{m_e^2c^3}{e\hbar}\approx4.41\times10^{13}\textrm{\ G},
    \label{eq:Brel}
\end{equation}
where $m_e$ is the electron mass, $\hbar$ is the Planck-Dirac constant and $e$ is the elementary electric charge.

Electrons are well approximated by an ideal Fermi gas. However, the presence of a magnetic field leads to the quantization of the electron motion perpendicular to the field into Landau-Rabi levels \citep{rabi1928freie,landau1930diamagnetismus}.
By ignoring thermal effects and the small electron anomalous magnetic moment, the electron energy density (with rest-mass excluded) and pressure are given by \citep{Chamel_2022}
\begin{equation}
    \rho _e=\frac{B_*m_ec^2}{\left ( 2\pi \right )^2\lambda_e^3}\sum_{\tilde{\nu}=0}^{\tilde{\nu}_{\textrm{max}}}g_{\tilde{\nu}} \left ( 1+2\tilde{\nu} B_* \right )\psi _+\left [ \frac{x_e\left ( \tilde{\nu} \right )}{\sqrt{1+2\tilde{\nu} B_*}} \right ]-n_e m_ec^2,
\end{equation}
\begin{equation}
    P_e=\frac{B_*m_ec^2}{\left ( 2\pi \right )^2\lambda_e^3}\sum_{\tilde{\nu}=0}^{\tilde{\nu}_{\textrm{max}}}g_{\tilde{\nu}}\left ( 1+2\tilde{\nu} B_* \right )\psi _-\left [ \frac{x_e\left ( \tilde{\nu} \right )}{\sqrt{1+2\tilde{\nu} B_*}} \right ],
\end{equation}
respectively, where $\lambda_e=\hbar/\left (m_ec \right )$ is the electron Compton wavelength, $g_{\tilde{\nu}}=1$ for $\tilde{\nu}=0$ and $g_{\tilde{\nu}}=2$ for $\tilde{\nu} \geq 1$,
\begin{equation}
    \psi _\pm \left ( x \right )=x\sqrt{1+x^2}\pm \textrm{ln}\left ( x+\sqrt{1+x^2} \right ),
\end{equation}
\begin{equation}
    x_e\left (\tilde{\nu}\right )=\sqrt{\gamma_e^2-1-2\tilde{\nu} B_*},
    \label{eq:xe}
\end{equation}
where $\gamma_e$ is the electron Fermi energy in the units of $m_ec^2$. $\tilde{\nu}_{\textrm{max}}$ is related to the electron number density $n_e$ by \cite{Chamel_2022}
\begin{equation}
    n _e=\frac{2B_*}{\left ( 2\pi \right)^2\lambda_e^3}\sum_{\tilde{\nu}=0}^{\tilde{\nu}_{\textrm{max}}}g_{\tilde{\nu}} x_e\left (\tilde{\nu} \right ).
\end{equation}
The index $\tilde{\nu}_{\textrm{max}}$ is the highest integer for which the combination inside the square root of Eq.~(\ref{eq:xe}) is larger than or equal to zero, i.e., 
\begin{equation}
    \tilde{\nu}_{\textrm{max}}=\left [ \frac{\gamma_e^2-1}{2B_*} \right ],
\end{equation}
where $\left [.\right ]$ denotes the integer part. 
According to the condition of charge neutrality, the average baryon number density is given by
\begin{equation}
    \bar{n}=\frac{A}{Z}n_e=An_N.
\end{equation}
For pointlike ions embedded in a uniform electron gas, the energy density associated to the 
electron-ion interactions is given by (see, e.g., Chapter 2 of \cite{ALWattsTEStrohmayer2007})
\begin{equation}
    \rho _L=C_M\left ( \frac{4\pi}{3} \right )^{1/3}e^2n_e^{4/3}Z^{2/3},
\end{equation}
where $C_M$ is the Madelung constant. The corresponding contribution to the pressure is therefore given by
\begin{equation}
    P_L=n_e^2\frac{d\left ( \rho _L/n_e \right )}{dn_e}=\frac{\rho _L}{3} .
\end{equation}
The total pressure of the Coulomb plasma is thus given by $P=P_e+P_L$, and the corresponding energy density is $\rho=\rho_N+\rho_e+\rho_L$. In this work, the Wigner-Seitz estimate $C_M=-9/10$ will be adopted for the Madelung constant \citep{Salpeter1954ElectronSA}.


The composition of the crust in thermodynamic equilibrium under the presence of magnetic field is determined by minimizing the Gibbs free energy per nucleon, which coincides with the baryon chemical potential 
(see, e.g., Appendix A in \cite{PhysRevD.92.023008}):
\begin{equation}
\begin{split}
    \mu_{c}=\frac{\rho+P}{\bar{n}}=&\frac{M'\left ( A,Z \right )c^2}{A}+\\ &\frac{Z}{A}m_ec^2\left [ \gamma_e-1+\frac{4}{3}C_M\left ( \frac{4\pi}{3}^{1/3}\alpha\lambda_e n_e^{1/3}Z^{2/3} \right ) \right ],
\end{split}
\end{equation}
where $\alpha$ is the fine-structure constant.
When descending in the crust region from a layer made of nuclei $\left (A_1,Z_1\right )$ to a denser layer made of nuclei $\left (A_2,Z_2\right )$, the pressure $P_{1\rightarrow2}$ associated to the transition is determined by 
the equilibrium condition
\begin{equation}
    \mu_c\left (A_1,Z_1,P_{1\rightarrow2}\right )=\mu_c\left (A_2,Z_2,P_{1\rightarrow2}\right ),
\end{equation}
which can be approximately written as \cite{Chamel_composition_2020}
\begin{equation}
    \gamma_e+C_M\alpha\lambda_e F\left ( Z_1,A_1;Z_2,A_2\right )\left (  \frac{4\pi n_e}{3}\right )^{1/3}=\gamma_e^{1\rightarrow 2},
\end{equation}

\begin{equation}
\begin{split}
    F&\left ( Z_1,A_1;Z_2,A_2\right )\equiv \\ 
    &\left ( \frac{4}{3}\frac{Z_1^{2/3}Z1}{A_1}-\frac{1}{3}\frac{Z_1^{2/3}Z_2}{A_2}-\frac{Z_2^{2/3}Z_2}{A_2} \right )\left ( \frac{Z_1}{A_1}-\frac{Z_2}{A_2} \right )^{-1},
\end{split}
\end{equation}

\begin{equation}
\begin{split}
    \gamma_e^{1\rightarrow 2}&\equiv 1\;\;+ \\
    &\left [ \frac{M'\left ( A_2,Z_2 \right )}{A_2m_e}-\frac{M'\left ( A_1,Z_1 \right )}{A_1m_e} \right ]\left ( \frac{Z_1}{A_1}-\frac{Z_2}{A_2} \right )^{-1}.
\end{split}
\end{equation}
For the detailed composition and nuclear parameters, we refer the reader to Tables 2 and 3 in 
\cite{Chamel_2022}.

\subsection{Inner crust}


The effects of Landau-Rabi quantization in the inner crust have been implemented within the fourth-order extended Thomas Fermi method with proton shell corrections added consistently via the Strutinsky integral \citep{Mutafchieva_2019}. 
By ignoring the neutron band-structure effects, the boundary between the inner and outer crustal region is determined by the condition $\mu_c=m_nc^2$, where $m_n$ is the mass of neutron \citep{PhysRevC.91.065801,PhysRevC.91.055803}. Nuclear clusters\footnote{We refrain here from referring to clusters in the inner crust as 'nuclei'. Indeed, those extremely neutron-rich clusters would immediately disintegrate if they were placed in vacuum. Their existence in a neutron star is only possible because of the degenerate neutron gas blocking neutron emission.} are assumed to be spherical and 
unaffected by the magnetic field. The Coulomb lattice is considered using the 
Wigner-Seitz approximation
and the nucleon density distributions in the Wigner-Seitz cell are parameterized as 
\begin{equation}
    n_q\left (r'\right )=n_{B,q}+n_{\Lambda,q}f_q\left( r'\right ),
\end{equation}
where $q=p(n)$ for protons (neutrons), $n_{B,q}$ are the background nucleon number densities (typically protons are bound inside clusters therefoer $n_{B,p}\approx 0$), $n_{\Lambda,q}$ are the nucleon number densities characterizing the clusters, $r'$ is the radial coordinate for the Wigner-Seitz cell and $f_q\left( r'\right )$ describes the spatial inhomogeneities. This function is given by
\begin{equation}
 f_q\left ( r' \right )= \left \{ 1+\exp \left[ \left( \frac{C_q-a_i}{r'-a_i} \right)^2-1 \right]
 \exp \left(  \frac{r'-C_q}{a_q}  \right) \right \}^{-1},
\end{equation}
in which
\begin{equation}
a_i=\left ( \frac{3}{4\pi n_N} \right )^{1/3},
\end{equation}
where $a_i$ is the ion-sphere radius, $C_q$ is the cluster radius defined as the half width at half maximum and $a_q$ is the diffuseness of the cluster surface. 
The numbers of neutrons and protons in the Wigner-Seitz cell are respectively determined by
\begin{equation}
    N=4\pi\int_{0}^{a_i}r'^2n_n\left ( r' \right )dr',
\end{equation}
\begin{equation}
    Z=4\pi\int_{0}^{a_i}r'^2n_p\left ( r' \right )dr'.
\end{equation}

The nuclear energy density functional BSk24 \citep{goriely2013further}, which underlies 
the nuclear mass model HFB-24 used for the outer crust, is employed to determine the EOS
of nuclear clusters and free neutrons. 
The energy per nucleon $E/A$ is minimized at fixed average baryon density $\bar{n}$, which 
is given by
\begin{equation}
    \frac{E}{A}=\frac{4\pi}{A}\int_{0}^{a_i}r'^2\varepsilon\left ( r' \right )dr',
\end{equation}
where $\varepsilon\left (r'\right )$ is the energy density and $A=N+Z$ is the total number of nucleons in the Wigner-Seitz cell \citep{2018MNRAS.481.2994P}, taking into account 
the contributions of magnetized electron Fermi gas as discussed in \cite{Mutafchieva_2019}.


\subsection{Core}

\begin{figure}
	\includegraphics[width=\columnwidth]{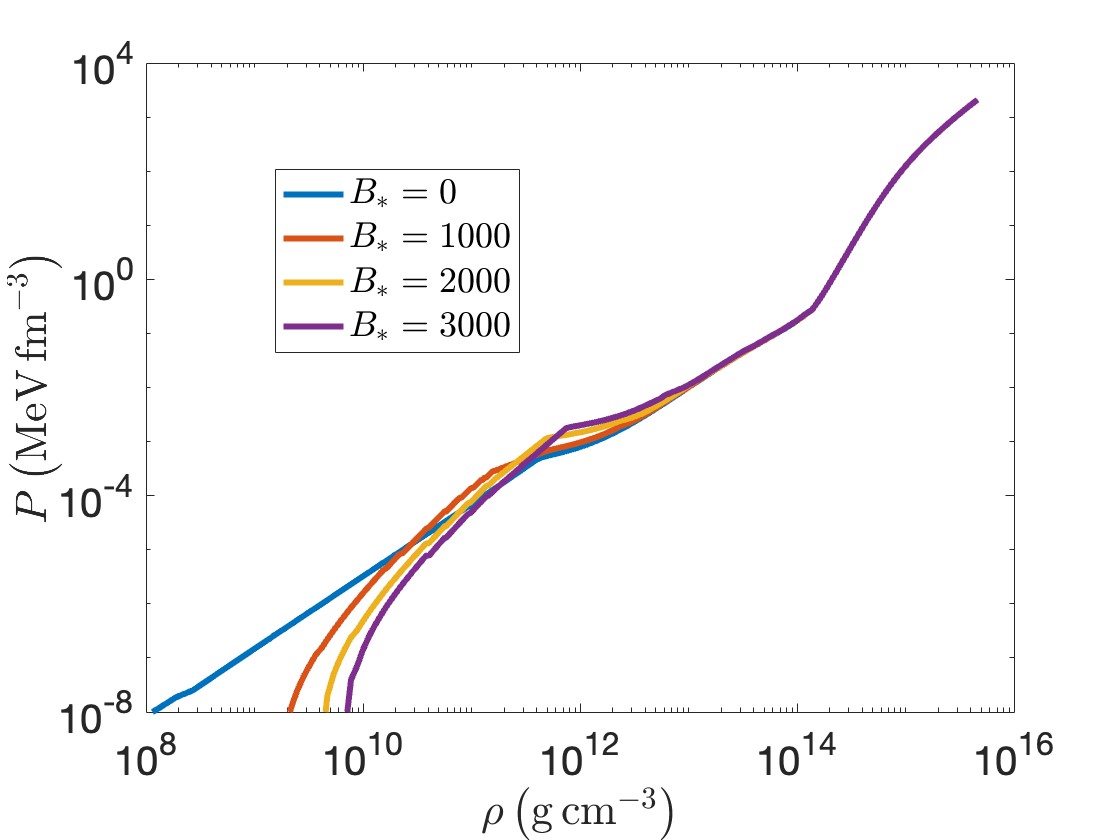}
    \caption{The pressure $P$ is plotted against the energy density $\rho$ for different values of normalized magnetic-field strength $B_* \equiv B/B_{\rm rel}$, where
    $B_{\rm rel}$ is defined in Eq.~(\ref{eq:Brel}). }
    \label{fig:EOS}
\end{figure}

The zero temperature unified EOS in beta equilibrium based on the same Brussels-Montreal energy-density functional BSk24 \cite{BSk24_core1} is employed for the dense fluid core.  
This is the same EOS for unmagnetized neutron stars and we refer the reader to \cite{BSk24_core2,2018MNRAS.481.2994P} for details. This EOS was shown to be consistent with astrophysical observations, including the gravitational-wave signal from GW 170817 and its electromagnetic counterpart \citep{PhysRevC.100.035801}.
As an illustration, Fig.~\ref{fig:EOS} plots the pressure $P$ against energy density $\rho$ for the complete EOS model at different normalized magnetic field strength $B_*$ for comparison. 
It should be a good approximation as the magnetic field 
has negligible effects on the EOS in the high density core for the range of field strength
considered in this work (see Sec.~\ref{sec:SELF-CONSISTENT STELLAR MODELS} for further discussion). 
Indeed, as can be seen in Fig.~\ref{fig:EOS}, the EOS of the crust matches smoothly that of the core.

\subsection{Shear modulus of the crust}
\label{sec:Shear modulus}

\begin{figure}
	\includegraphics[width=\columnwidth]{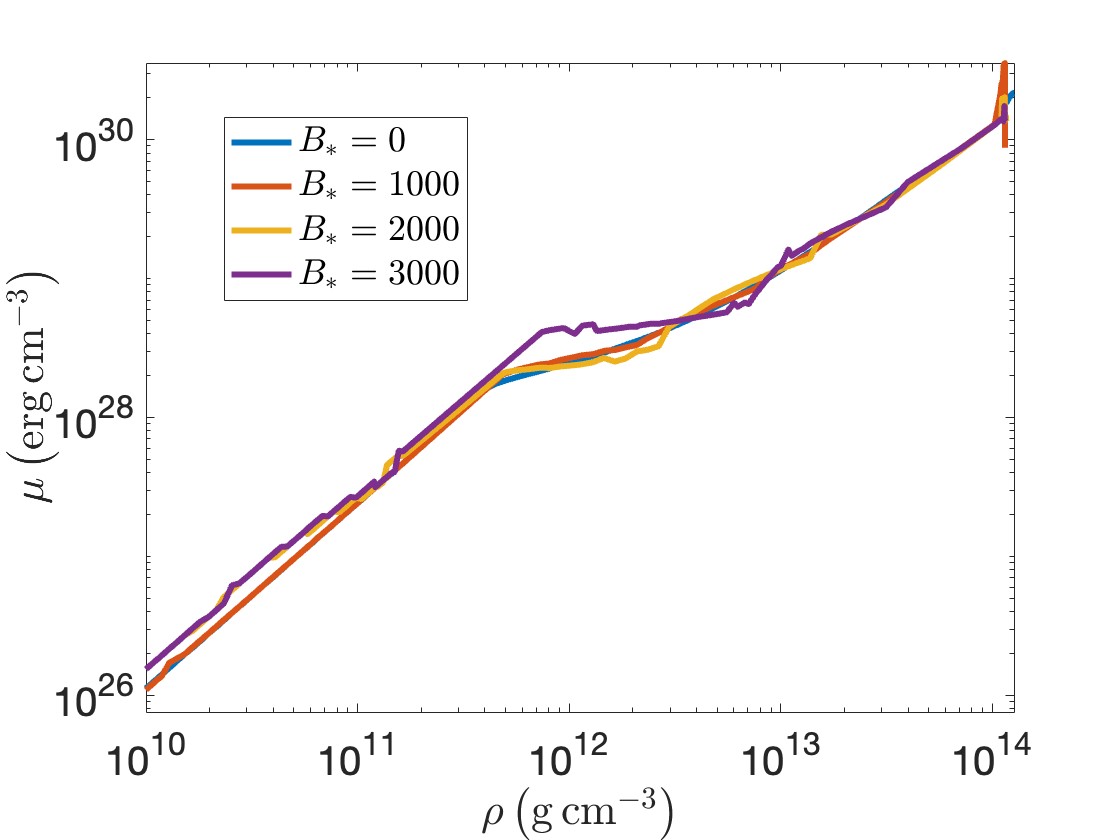}
    \caption{The shear modulus $\mu$ as a function of the energy density $\rho$ for different values of normalized magnetic-field strength $B_* \equiv B/B_{\rm rel}$, where
    $B_{\rm rel}$ is defined in Eq.~(\ref{eq:Brel}).}
    \label{fig:shear}
\end{figure}

The elastic property of the crystallized matter formed by the ions in the neutron star crust treated as an isotropic polycrystal can be characterized by an effective shear modulus $\mu$. 
Since the shear modulus in the crust is responsible for the restoring force of torsional oscillations, the frequencies of torsional oscillation modes strongly depend on the shear modulus. 
To compute the torsional modes, an estimation for the shear modulus $\mu$ is required.
For practical applications in the neutron star crust, the model of isotropic Coulomb 
solid \cite{OgataIchimaru1990} is considered and the shear modulus (in CGS system of units) is given by
\begin{equation}
\mu=0.1194\:n_N\frac{Z^2e^2}{a_i}.
\end{equation}
In Fig.~\ref{fig:shear}, we plot $\mu$ against $\rho$ for our EOS with different values of $B_*$.

\section{MAGNETIZED STELLAR MODELS}
\label{sec:SELF-CONSISTENT STELLAR MODELS}

In standard calculations of the torsional modes of magnetized neutron stars (e.g., \citep{Sotani2007}), 
one needs to first obtain the spherically symmetric background star by solving the TOV equation. A magnetic field configuration is then imposed and solved for the equilibrium stellar model as discussed in Sec.~\ref{sec:MIXED POLOIDAL-TOROIDAL MAGNETIC FIELDS}.
However, when the EOS of the crust depends on the magnetic field, so will the structure of the crust.
The standard approach is thus not applicable here.

\begin{figure}
	\includegraphics[width=\columnwidth]{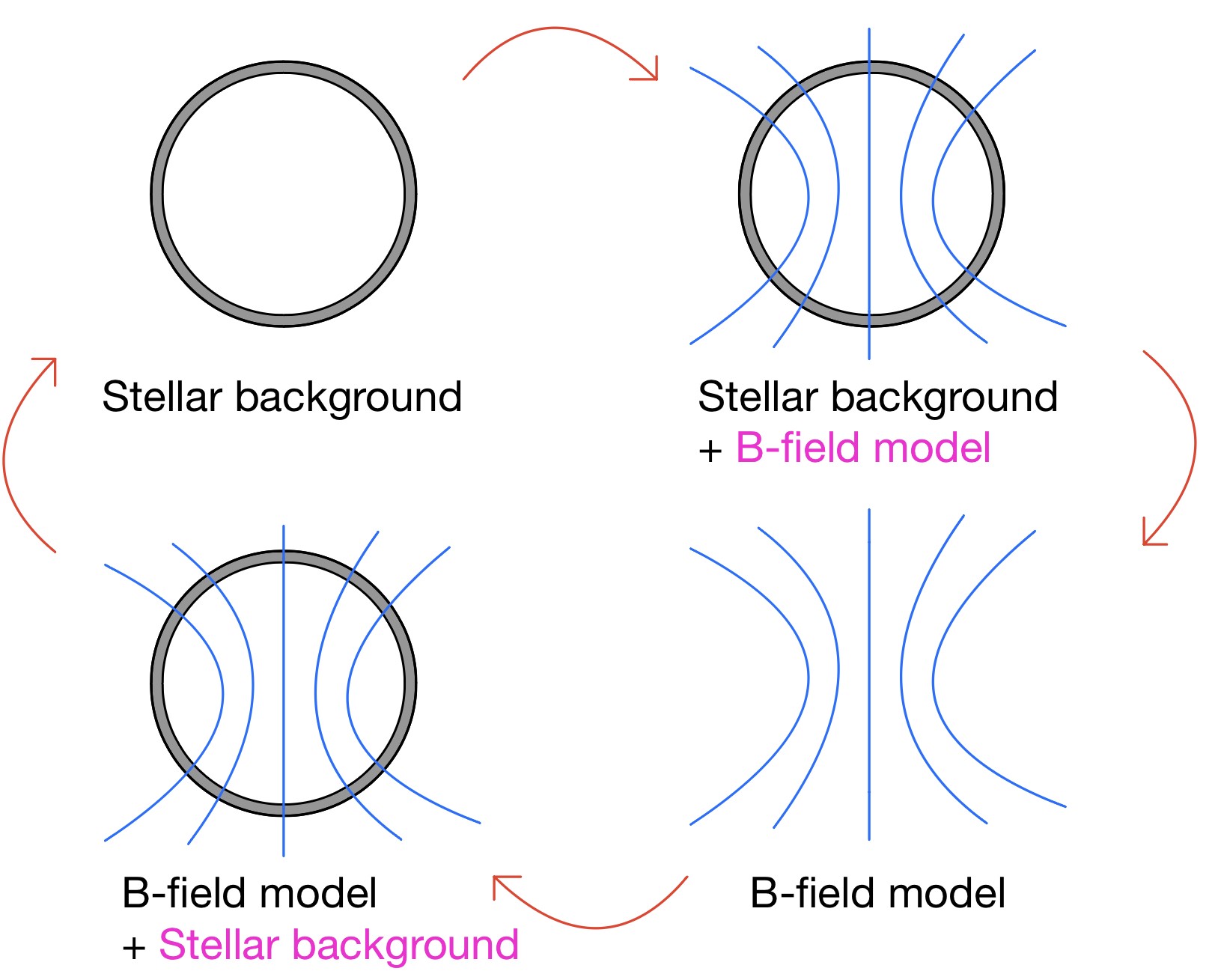}
    \caption{Graphical illustration of the iterative method for the construction of a magnetized stellar model starting from the upper-left picture. The words in black indicate the information obtained from the previous iteration while those in pink indicate the information to be determined in the current iteration 
    (see text for more details).  }
    \label{fig:self consistent loop}
\end{figure}

In order to construct a magnetized equilibrium stellar model that includes the effects of magnetic field on the EOS, we first construct a stellar model without magnetic field and then impose a 
magnetic field configuration on it just like the standard approach outlined above. 
Next, with this magnetic field configuration, a new stellar background model is calculated by solving the TOV equation with a magnetic-field-dependent EOS. After that, a new magnetic field configuration is constructed using the new stellar background. This process is repeated until the properties of the star (e.g., the radius $R$ and the crust thickness $\delta r$) converge to some values. When this is achieved, we declare that a magnetized stellar model is obtained. Our method is summarized by the iteration loop presented in Fig.~\ref{fig:self consistent loop}.

As an illustration, Table~\ref{tab:converge} presents the convergence of $R$ and $\delta r$ after each iteration for a 1.4 $M_\odot$ magnetized neutron star with a normal fluid core and a dipole magnetic field having the magnetic field strength of $4\times10^{16}$ G at the pole of the surface. The convergence is achieved when the new $\delta r$ only differs from the previous $\delta r$ by less than 0.1\%. 
It is seen that the values of $R$ and $\delta r$ converge rapidly with a few number of iteration
cycles only. However, the number of iterations will be greater for stronger magnetic field strengths and more complex magnetic field configurations.

\begin{table}
	\centering
	\caption{The values of $M$, $R$ and $\delta r$ for a $1.4 M_\odot$ neutron star model with a normal
 fluid core during the iterative calculation of a magnetized stellar model. The magnetic-field configuration is purely dipole and the field strength at the pole of the surface is $4\times10^{16}$ G. }
	\label{tab:converge}
	\begin{tabular}{lccr} 
		\hline
        No. of cycle 
        iteration & $M/M_\odot$ & $R$ (km) & $\delta r$ (km)\\
		\hline
		0 & 1.40 & 12.4360 & 0.8722\\
		1 & 1.40 & 12.5426 & 0.9765\\
        2 & 1.40 & 12.5448& 0.9786\\
        3 & 1.40 & 12.5449& 0.9787\\
        \hline

	\end{tabular}
\end{table}

In our study, two approximations are made to simplify the problem. 
Firstly, we assume that the magnetic field has negligible effect on the structure of the fluid core, and hence the fluid core of the star is fixed during the iteration. 
This should be a good approximation as the EOS at the high-density range relevant
to the core is insensitive to the magnetic field for the field strength being
considered in this work \citep{Mutafchieva_2019}. 
It has also been pointed out \cite{Chatterjee_2021} that the EOS in the core may be affected by the magnetic field only for field strength of the order of $10^{18}$ G, which is beyond our range of consideration. 

Secondly, when we solve for the spherically symmetric model using the TOV equation with a magnetic-field-dependent EOS, we make use of the root mean square (r.m.s.) values of the magnetic field $B\left (r, \theta \right )$ on any shell of radius $r$. 
For a given magnetic field configuration, the magnitude of magnetic field in general depends
on both the radial and angular positions. In principle, the stellar structure will no longer be spherically symmetric, though the deviation is expected to be very small for the field strength being considered in this work. 
By using the r.m.s. values of $B\left (r, \theta \right )$, the background stellar model is kept spherically symmetric, but its radius and crust thickness 
will in general be different from those obtained 
without considering the effects of the magnetic field on the EOS.

According to our simplified magnetic field models, $B_\phi$ is discontinuous at the stellar surface when $\xi > 0$, which is unphysical as already noted in Sec.~\ref{sec:MIXED POLOIDAL-TOROIDAL MAGNETIC FIELDS}. 
During the iterative process of constructing a new equilibrium model that incorporates the effects of 
magnetic field on the EOS, the stellar radius increases slightly. 
As a result, a thin vacuum layer just outside the star in the previous iteration step becomes 
filled with matter at the current step. 
To account for this effect, we extrapolate the magnetic field within the matter but not outside the star for the next iteration. The continuity of the magnetic field inside the star is maintained throughout this 
process.
The method of extrapolating $B_\phi$ is to keep the corresponding value of $\xi$ when the magnitude of 
$B_\phi$ is being calculated using the expression shown in Eq.~(\ref{eq:B_component}). 
At the end of the iteration when a new equilibrium model is obtained, the magnetic field still exhibits the same characteristic, where $B_\phi$ remains discontinuous at the stellar surface for $\xi > 0$, as discussed in the standard treatment outlined in Sec.~\ref{sec:MIXED POLOIDAL-TOROIDAL MAGNETIC FIELDS}.


\begin{figure}
	\includegraphics[width=0.95\columnwidth]{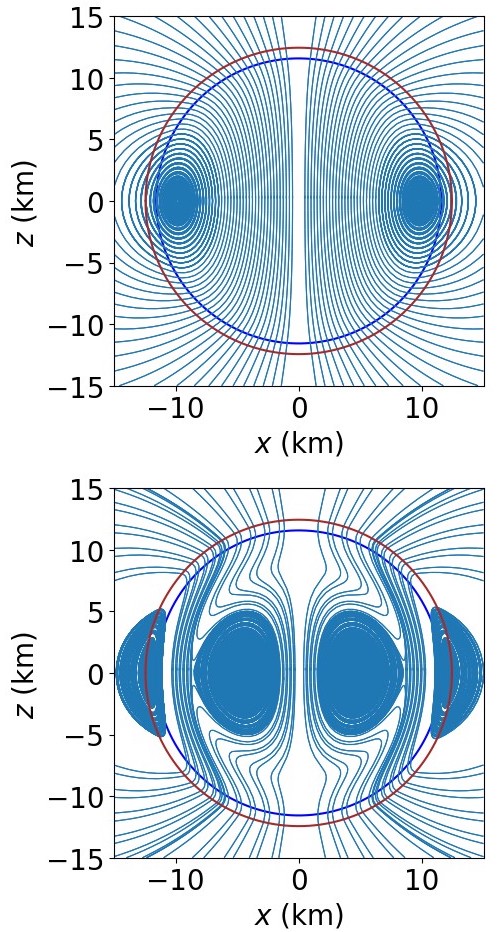}
    \caption{Magnetic field lines on the meridional plane for 1.4 $M_\odot$ magnetized neutron stars with a 
    normal fluid core. The upper (lower) panel shows the case for $\xi=0$ (0.45). In each panel, the red and blue circles indicate the stellar surface and the crust-core interface, respectively. }
    \label{fig:combined_b_field_configuration}
\end{figure}

As an illustration, Fig.~\ref{fig:combined_b_field_configuration} shows different magnetic field configurations on the meridional plane for 1.4 $M_\odot$ neutron stars with a normal fluid core. The upper (lower) panel
shows the case for $\xi=0$ (0.45). In each panel, the red and blue circles indicate the stellar surface
and the crust-core interface, respectively. One can see that the complexity of the magnetic field configuration
increases with $\xi$. 

In the following, the variable $B$ will be used to represent the strength of magnetic field at the pole of the stellar surface. 

\section{TORSIONAL OSCILLATIONS}
\label{sec:TORSIONAL OSCILLATIONS}

\subsection{General-relativistic pulsation equations}

For the torsional oscillation modes, the only perturbed fluid variable is the azimuthal component of the perturbed 4-velocity \cite{SchumakerThorne1983} 
\begin{equation}
\label{eq:azimutha component}
    \delta u^\phi= \frac{e^{-\nu} }{
\sin\,\theta}\frac{\partial Y(r,t)}{\partial t}\frac{d P_\ell (\cos\,\theta)}{d \theta},
\end{equation}
where $P_\ell(\cos\,\theta)$ is the Legendre polynomial of order $\ell$. 
A harmonic time-dependence is assumed such that the perturbation function $Y(r,t) = Y(r)e^{i \sigma t}$, where 
$\sigma$ is the angular mode frequency.

In the relativistic Cowling approximation where the metric perturbations are neglected, the equations for determining the torsional oscillation modes can be obtained (see \cite{Messios_2001,Sotani2007} for details) by linearizing the fluid equation and the magnetic induction equation, which is derived from the homogeneous Maxwell's equations with the ideal magnetohydrodynamical approximations such that the electric field 4-vector vanishes.

In the following, we briefly summarize the main perturbation equations studied in \cite{Sotani2007} relevant
to our discussion. The perturbed magnetic induction equation is given by
\begin{equation}
\label{eq:magnetic induction equation}
    \delta\left \{ \left ( u^\alpha H^\beta - H^\alpha u^\beta \right )_{;\: \beta} \right \}=0 , 
\end{equation}
and the perturbed fluid equation is 
\begin{equation}
\label{eq:momentum-conservation equation}
    \delta\left ( \left [ \delta^j_\alpha+u^ju_\alpha \right ]T^{\alpha \beta}_{\ \ \ \ \  ;\ \beta} \right )=0, 
\end{equation}
where the spatial index $j=(r,\theta,\phi)$; $T^{\alpha \beta}$ is the stress-energy tensor for the background 
magnetized neutron star 
\begin{equation}
\label{eq:energy momentum tensor}
    T^{\alpha \beta}=\left ( P+\rho+H^2 \right )u^\alpha u^\beta+\left ( P+\frac{H^2}{2} \right )g^{\alpha \beta}-H^\alpha H^\beta, 
\end{equation}
where $g^{\alpha \beta}$ is the (inverse) spacetime metric. 
We assume the background equilibrium star is described by a perfect fluid under zero strain, and thus the stress is given by the isotropic pressure. The shear stress due to elasticity only contributes at the perturbative level in
Eqs.~(\ref{eq:magnetic induction equation}) and (\ref{eq:momentum-conservation equation}) after perturbing $T^{\alpha \beta}$ (see \cite{Sotani2007} for details).

When attempting to separate variables of the system of equations using Eq.~(\ref{eq:azimutha component}), 
it has been suggested by \citep{Sotani2007} that the presence of a mixed poloidal-toroidal magnetic field causes the eigenfunction of a single mode, which would initially correspond to a single value of $\ell=\ell_0$ in the non-magnetized case, to couple with eigenfunctions of other values of $\ell$, namely $\ell_0-2$ and $\ell_0+2$. However, it has been found that this coupling is unimportant for calculating the lowest-order mode. 
We refer the reader to Fig. 3 in \citep{Sotani_2008_phi} for a comparison of the frequencies of fundamental torsional modes of $\ell = 2$ with and without taking into account the $\ell \pm 2$ couplings.
To simplify the problem and since we will only focus here on the fundamental torsional modes of $\ell=2$, 
the $\ell \pm 2$ couplings will be ignored in the following. 
The required final pulsation equation obtained from combining 
Eqs.~(\ref{eq:magnetic induction equation}) and (\ref{eq:momentum-conservation equation}) is given by \cite{Sotani2007}
\begin{widetext}
\begin{eqnarray} 
\label{eq:pulsation equation}
&&\frac{d^2Y }{d r^2} + \left  [ \left ( 1+2\lambda_1 \right )\frac{a_1^2}{\mu\pi r^4} \right ] \frac{d^2Y }{d r^2} + \left \{  \left ( \frac{4}{r}+\frac{d\nu }{d r}-\frac{d\lambda }{d r}+\frac{1}{\mu}\frac{d\mu }{d r} \right ) 
+\left ( 1+2\lambda_1 \right )\frac{a_1}{\mu \pi r^4}\left [ \left ( \frac{d\nu }{d r}-\frac{d\lambda }{d r} \right )a_1+2\frac{da_1 }{d r} \right ]   \right \} \frac{dY }{d r}   \cr
&&\cr
&& + \left. \right. e^{2\lambda - 2\nu} \left [   \left ( \rho + P \right )\frac{\sigma^2}{\mu}-e^{2\nu}\frac{(\ell-1)(\ell+2)}{r^2}    \right ] Y
+ \left \{  \left ( 1+2\lambda_1 \right )\frac{a_1^2}{\mu\pi r^4} e^{2\lambda - 2\nu}\sigma^2-\frac{\lambda_1}{2\mu\pi r^2}\left (\frac{da_1 }{d r}\right )^2\sigma^2e^{-2\nu} \right. \cr
&&\cr
&& + \left.  \left (\ell-1 \right )\left (\ell+2\right )\frac{\lambda_1}{2\mu\pi r^4}\left (\frac{da_1 }{d r}\right )^2
+\left ( 2+5\lambda_1 \right )\frac{a_1}{2\mu \pi r^4}\left [ \left ( \frac{d\nu }{d r}-\frac{d\lambda }{d r} \right )\frac{da_1 }{d r}+\frac{d^2a_1 }{d r^2} \right ]   \right \}  Y =0 ,
\end{eqnarray}
where $\lambda_1=-\frac{\ell(\ell+1)}{(2\ell-1)(2\ell+3)}\:$. Using Eq.~(\ref{eq:solve b-field}) and the variables
\begin{equation}
\begin{split}
Y_1\equiv Yr^{1-\ell},\; \; \; \; \; \; \; \; \; \; \; \; \; \; \; \; \; \; \; \; \; \; 
Y_2\equiv \left [ \mu+\left ( 1+2\lambda_1 \right )\frac{a_1^2}{\pi r^4} \right ]e^{\nu-\lambda}r^{2-\ell}\frac{dY }{d r},
\end{split}
\end{equation}
Eq.~(\ref{eq:pulsation equation}) then becomes two equations in terms of $Y_1$ and $Y_2$:
\begin{eqnarray}
\frac{dY_1 }{d r} &=& -\frac{\ell-1}{r}\:Y_1+\frac{\pi r^3}{\pi r^4\mu+(1+2\lambda_1)a_1^2}\:e^{-\nu+\lambda}\:Y_2,   \\
&&\cr
\frac{dY_2 }{d r} &=& -\left \{  \left [ \rho+P+\left ( 1+2\lambda_1 \right )\frac{a_1^2}{\pi r^4} -\frac{\lambda_1e^{-2\lambda}}{2\pi r^2}\left (\frac{da_1 }{d r}\right )^2 \right ] 
\sigma^2 r e^{2\left ( \lambda-\nu \right )} -  ( \ell-1 ) ( \ell+2 ) \left [ \frac{\mu e^{2\lambda}}{r}-\frac{\lambda_1}{2\pi r^3} \left (\frac{da_1 }{d r}\right )^2 \right ] \right. \cr
&&\cr
&& + \left. ( 2+5 \lambda_1 )\frac{a_1e^{2\lambda}}{\pi r^3} \left [ \frac{a_1}{r^2}-\frac{\xi^2 e^{-2\nu}a_1}{2}-2\pi f_0\left ( \rho+P \right )r^2 \right ] \right \} e^{\nu-\lambda}Y_1 - 
\frac{\ell+2}{r}Y_2  .
\label{eq:Y_2_equation}
\end{eqnarray}
This is the first-order system of equations we solved for the real eigenvalues $\sigma$.
\end{widetext}

\subsection{Boundary conditions}

Boundary conditions are needed to be imposed to solve the above system of equations for the mode frequencies. 
As torsional modes are confined to the crust region, a zero traction condition at the core-crust interface and a zero torque condition at the stellar surface implies that $Y_2 = 0$ at both $r = R-\delta r$ and $r = R$ \cite{Sotani2007}. 
However, it should be pointed out that in reality one should also consider the coupling of the crust
to the fluid core \cite{Sotani_2008_phi}. In this work, we ignore the coupling with the fluid core to demonstrate the effects of magnetic-field-dependent EOS on the spectrum of torsional modes.


\subsection{Code tests}
\label{sec:CODE TESTS}

Before studying the effects of Landau-Rabi quantization, we first test the torsional mode frequencies 
$_\ell f_n$ produced by our code with the results presented in \cite{Sotani2007,Sotani_global,Sotani_2008_phi}, which do not consider the effects of magnetic field on the EOS. 

Firstly, the mode frequencies in the non-magnetized limit $\bar{_\ell f_n}$ are tested. Our results were compared with those in \cite{Sotani2007} for $\bar{_\ell f_n}$ with $n = 0,1$ (which is the number of nodes along the coordinate $r$, from the center to stellar surface) and $\ell=2,3,4,...,10$ for 
their stellar models A+DH$_{1.4/1.6}$, WFF3+DH$_{1.4/1.6}$, A+NV$_{1.4/1.6}$ and WFF3+NV$_{1.4/1.6}$. 
For all the cases, we find that our mode frequencies differ from those reported in \cite{Sotani2007} 
by only about 1\%, which is much better than similar code tests reported in \cite{de_Souza_2019}. 
The deviations could be due to different numerical treatments of the tabulated EOS data in different 
codes. 

The coefficients $_\ell \alpha_n$ appeared in the following fitting formula proposed 
in \cite{Sotani2007}
\begin{equation}
\label{eq:sotani_fitting}
\frac{_\ell f_n}{\bar{_\ell f_n}}=\sqrt{1+ {_\ell \alpha_n} \left ( \frac{B}{B_\mu} \right )^2}  ,
\end{equation}
for purely dipolar magnetic configurations are also tested with $\ell=2$ and $n = 0,1$, where $B_\mu=4\times 10^{15}$ G. 
We again find that the values of $_\ell\alpha_n$ obtained for the same stellar models tested above differ from the 
results reported in \cite{Sotani2007} by only about 1\%. 

We then tested our mode frequencies $_2f_0$ with mixed poloidal-toroidal magnetic fields. In this part, no numerical data 
are given in \cite{Sotani_2008_phi}. So we compared our data with the results extracted digitally from the figures presented in \cite{Sotani_2008_phi}. The results are presented in Appendix~\ref{sec:Code test figure}.

\section{STRUCTURES OF NEUTRON STARS}
\label{sec:STRUCTURES OF NEUTRON STARS}

As mentioned above, having the magnetic-field-dependent EOS, the structure of a neutron star in general
varies with the strength and configuration of magnetic field. 
In this section, we will compare the differences in the stellar structures between 
models constructed with and without the effects of magnetic field on the EOS. 
In particular, we focus on the relative differences in $\delta r$ defined by 
\begin{equation}
\Delta_{\delta r}=\frac{\delta r' - \delta r^{\circ}}{\delta r^{\circ}},
\label{eq:Delta_dr}
\end{equation}
where $\delta r'$ is the thickness of the crust after considering the effects of 
magnetic-field-dependent EOS; $\delta r^{\circ}$ is the original thickness without the effects of magnetic field. 

We shall focus on 1.4 $M_\odot$ neutron stars as our canonical models in the study. For our chosen BSk24 EOS, a $1.4 M_\odot$ neutron star has crust thickness $\delta r^{\circ} = 0.8722$ km and radius $R^{\circ} = 12.4360$ km in the non-magnetized limit. A more massive $2.0 M_\odot$ model will be considered later in this work.

\subsection{Normal fluid cores}

\begin{figure}
	\includegraphics[width=\columnwidth]{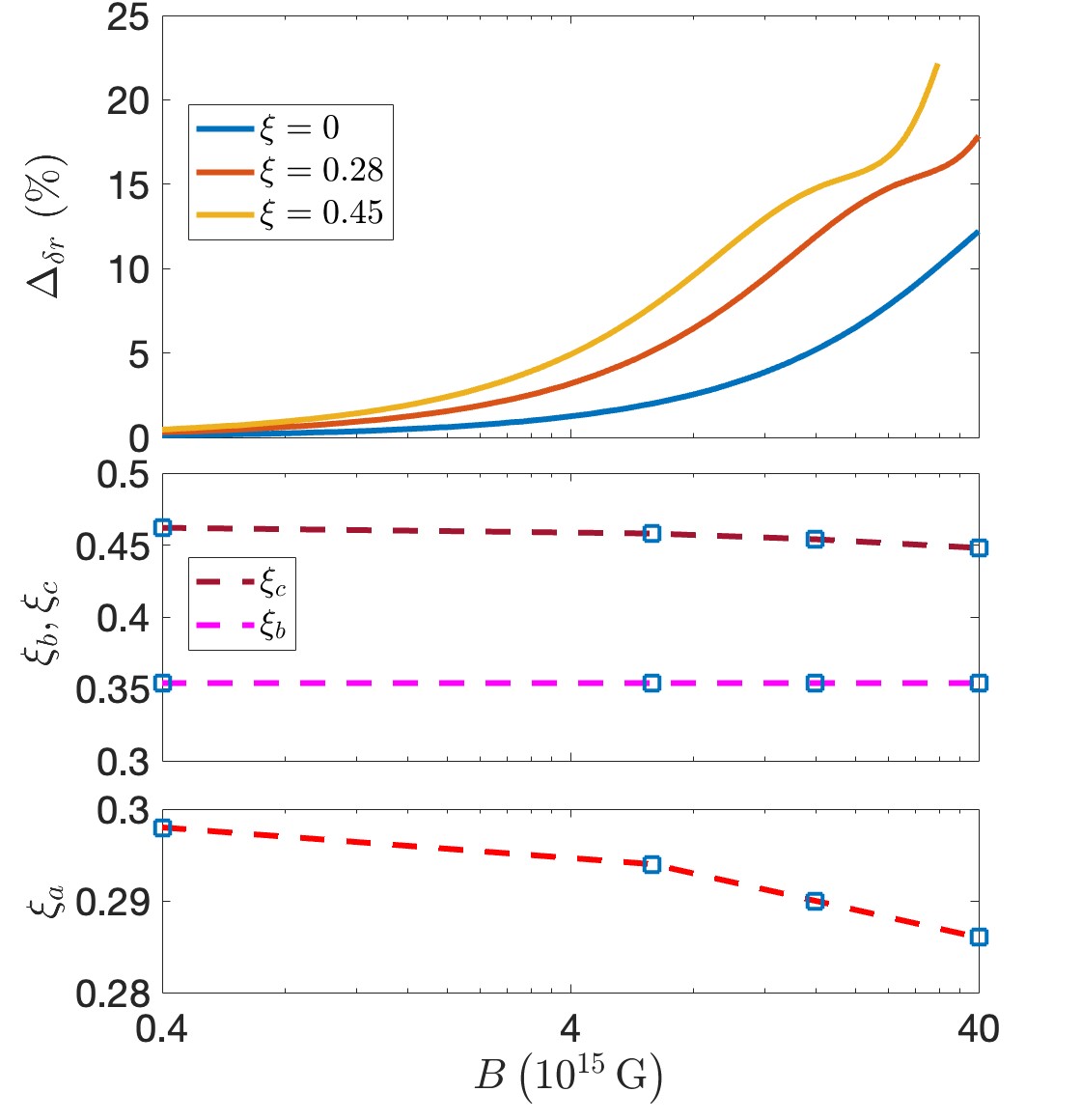}
    \caption{Upper panel: $\Delta_{\delta r}$ as a function of $B$ for $\xi = 0$, 0.28 and 0.45. Middle panel: $\xi_b$ and $\xi_c$ as a function of $B$. Lower panel: $\xi_a$ as a function of $B$. Here $B$ is the 
    magnetic-field strength at the pole of a magnetized neutron star, which 
    is assumed to have a normal fluid core. }
    \label{fig:combine_structure_normal}
\end{figure}

We first consider the case for normal fluid core and plot $\Delta_{\delta r}$ against the magnetic-field 
strength $B$ at the pole in the upper panel of Fig.~\ref{fig:combine_structure_normal}. In the figure, 
three different field configurations $\xi=0$, 0.28, and 0.45 are considered, where $\xi=0$ represents
the purely dipole case. 
It can be seen that $\Delta_{\delta r}$ in general increases with $B$ and $\xi$. Even for a purely dipole field, $\Delta_{\delta r}$ can exceed 10\% when the strength of magnetic field is of the order of $10^{16}$ G. 
For more complex magnetic field configurations with larger values of $\xi$, $\Delta_{\delta r}$ can even exceed 20\% when $B$ reaches over $10^{16}$ G. 
The strength of the magnetic field in the crustal region generally increases with $\xi$ (i.e., a stronger toroidal component) and hence leads to greater effects on the stellar structure via the EOS.

Increasing the crust thickness $\delta r$ also affects the allowed parameter space for 
physical magnetic-field configurations mentioned in Sec.~\ref{sec:MIXED POLOIDAL-TOROIDAL MAGNETIC FIELDS}. 
In the middle and bottom panels of Fig.~\ref{fig:combine_structure_normal}, we show how the parameters 
$\xi_a$, $\xi_b$, and $\xi_c$ defined in Eq.~(\ref{eq:Bfield_parameters}) change with the field
strength $B$.  
It is seen that $\xi_b$ and $\xi_c$ remain essentially unchanged. However, $\xi_a$ decreases 
more significantly as $B$ increases. As a result, the gap between the two physical ranges I and II defined 
in Eq.~(\ref{eq:Bfield_parameters}) increases with $B$. It should be noted that as the parameter 
$\xi$ increases towards the limiting value $\xi_a$, the field configuration in general would become more complex.  
In other words, for physical magnetic-field configurations defined in the physical range I, increasing the 
field strength $B$ would effectively increase the complexity of the field configuration for a fixed value of 
$\xi$. 
Similar effect also exists in the physical range II as it can be seen that its upper limit $\xi_c$ also decreases, though not as significant as $\xi_a$, when $B$ increases towards $4\times 10^{16}$ G.
The consequences of this important effect will be discussed in Sec.~\ref{sec:MODE FREQUENCIES}.

In order to understand why $\Delta_{\delta r}$ is positive and increases with the strength of the magnetic  
field, we first note that, for the same value of pressure $P$, increasing the field strength will result in a smaller value of energy density $\rho$ near the transition between the inner and outer crustal 
regions, which happens at the density range between about $5\times 10^{11} \ {\rm g \ cm}^{-3}$ and 
$10^{12} \ {\rm g \ cm}^{-3}$ for different field strengths shown in Fig.~\ref{fig:EOS}. 
According to Eq.~(\ref{eq:tov p}), a smaller value of $\rho$ leads to a smaller value of $|dP/dr|$, and hence a flatten density profile which can extend further away until it reaches the surface.

\begin{figure}
	\includegraphics[width=\columnwidth]{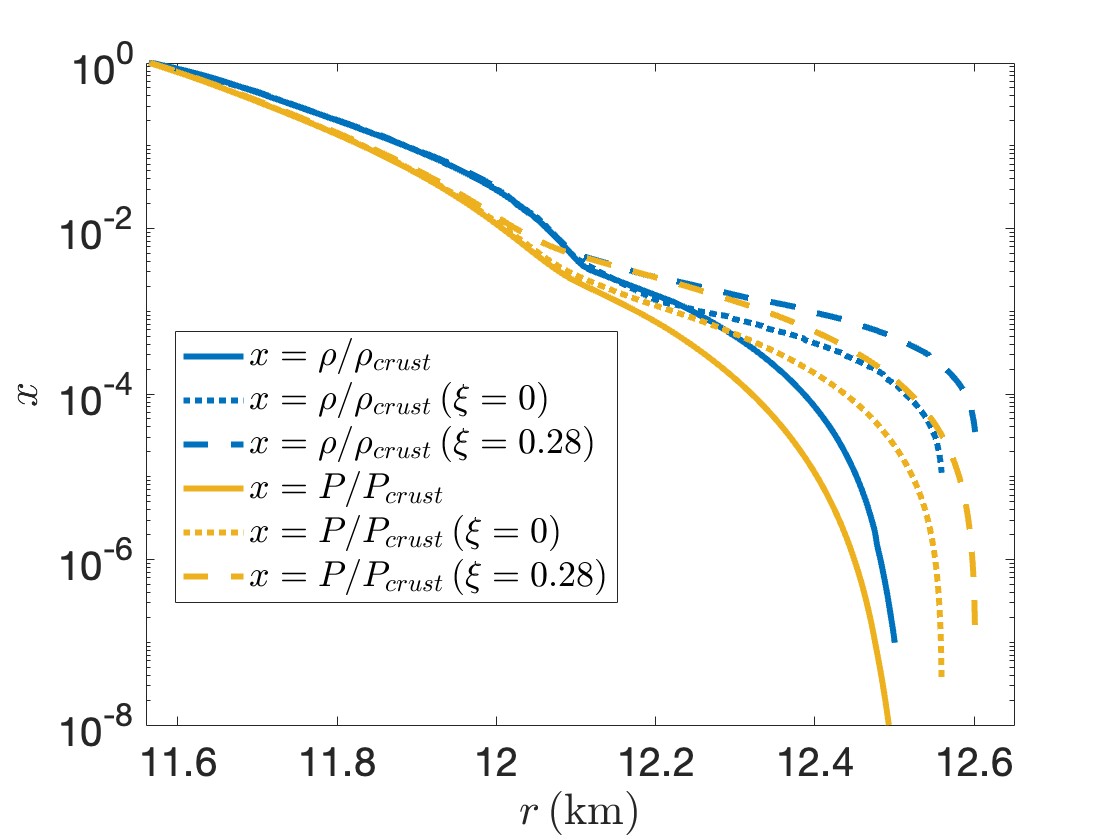}
    \caption{The profiles of energy density $\rho$ and pressure $P$ normalized by their own values 
    $\rho_{\rm crust}$ and $P_{\rm crust}$, respectively, at the base of the crust. 
    The solid lines represent the profiles for an unmagnetized neutron star. The dotted and dashed lines
    are the profiles for magnetized stars with the same magnetic field strength $B=4\times 10^{16}$ G, but
    with different field configurations $\xi=0$ and 0.28.     
    }
    \label{fig:parameter}
\end{figure}

To illustrate this phenomenon, we show the normalized energy-density profiles (blue lines) and pressure profiles (yellow lines) in the crustal region for three different star models in Fig.~\ref{fig:parameter}. The solid lines represent the profiles for a unmagnetized neutron star. The dotted lines are the profiles for a 
magnetized star with a purely dipole field ($\xi=0$) and $B=4\times 10^{16}$ G, which branch off from 
the solid lines leading to a thicker crust. 
However, the dotted lines in fact drop more rapidly near the stellar surface comparing to the solid lines. 
This can be understood from Fig.~\ref{fig:EOS} that increasing the field strength for a fixed pressure 
leads to a higher density at the low density region, and hence a steeper $|dP/dr|$ near the surface. 
One can also see from Fig.~\ref{fig:parameter} that, for the same field strength $B$ at the pole, increasing 
the value of $\xi$ from 0 to 0.28 (dashed lines) can further increase the thickness of the crust.

\subsection{Superconducting cores}

\begin{figure}
	\includegraphics[width=\columnwidth]{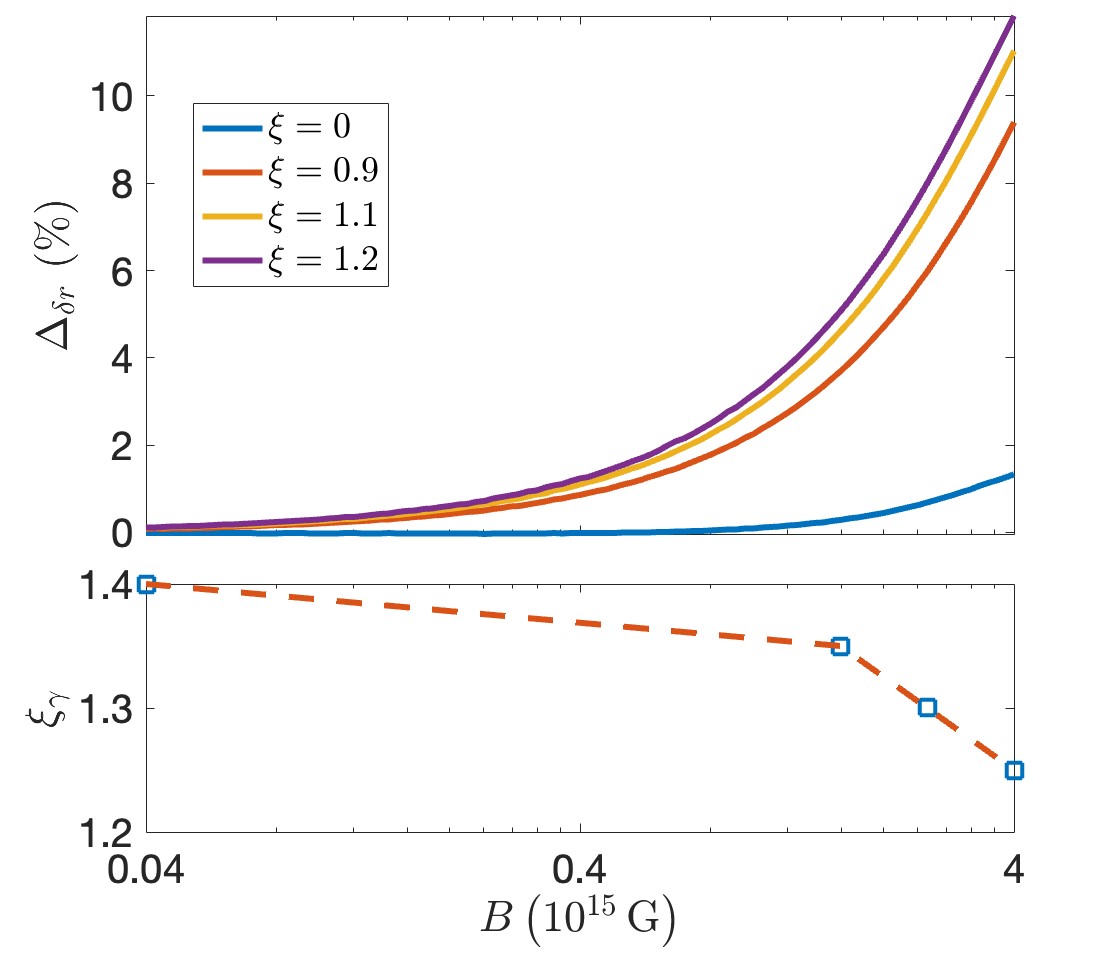}
    \caption{Upper panel: $\Delta_{\delta r}$ as a function of $B$ for $\xi = 0$, 0.9, 1.1 and 1.2. Lower panel: $\xi_\gamma$ as a function of $B$. The neutron star is assumed to have a superconducting core and 
    the magnetic field is confined in the crust. }
    \label{fig:combine_structure_superfluid}
\end{figure}

For neutron stars with a superconducting core, $\Delta_{\delta r}$ also increases with $B$ and $\xi$ as shown 
in the upper panel of Fig.~\ref{fig:combine_structure_superfluid}. 
However, $\Delta_{\delta r}$ now becomes more sensitive to the field strength comparing to the case of normal 
fluid core. 
For neutron stars with a normal fluid core, $\Delta_{\delta r}$ increases to about 5\% for $\xi=0.45$ and $B=4\times 10^{15}$ G as shown in Fig.~\ref{fig:combine_structure_normal}. 
For the same value of $B$, superconducting-core models can reach up to 10\% when $\xi \approx 1$. 
The bottom panel of Fig.~\ref{fig:combine_structure_superfluid} shows that the parameter 
$\xi_\gamma$, which defines the physical range of parameter in Eq.~(\ref{eq:Bfield_xi_gamma}), 
also decreases with increasing $B$.


\section{MODE FREQUENCIES}
\label{sec:MODE FREQUENCIES}

As mentioned above, the torsional oscillation modes $_\ell f_n$ for neutron stars with mixed poloidal-toroidal magnetic fields, but without considering the effect of Landau-Rabi quantization, have been studied \citep{de_Souza_2019}.  
In this work, we only focus on the $\ell =2$ fundamental torsional oscillation modes and investigate the changes in the mode frequencies when the effect of magnetic field on the EOS is considered. We define the relative differences in mode frequencies by

\begin{equation}
\Delta_f=\frac{{f}'-f^\circ}{f^\circ}, 
\label{eq:delta_definition}
\end{equation}
where $f'$ and $f^{\circ}$ are the frequencies calculated with and without the effect of magnetic field on 
the EOS, respectively.

\subsection{Normal fluid cores}

\begin{figure}
	\includegraphics[width=\columnwidth]{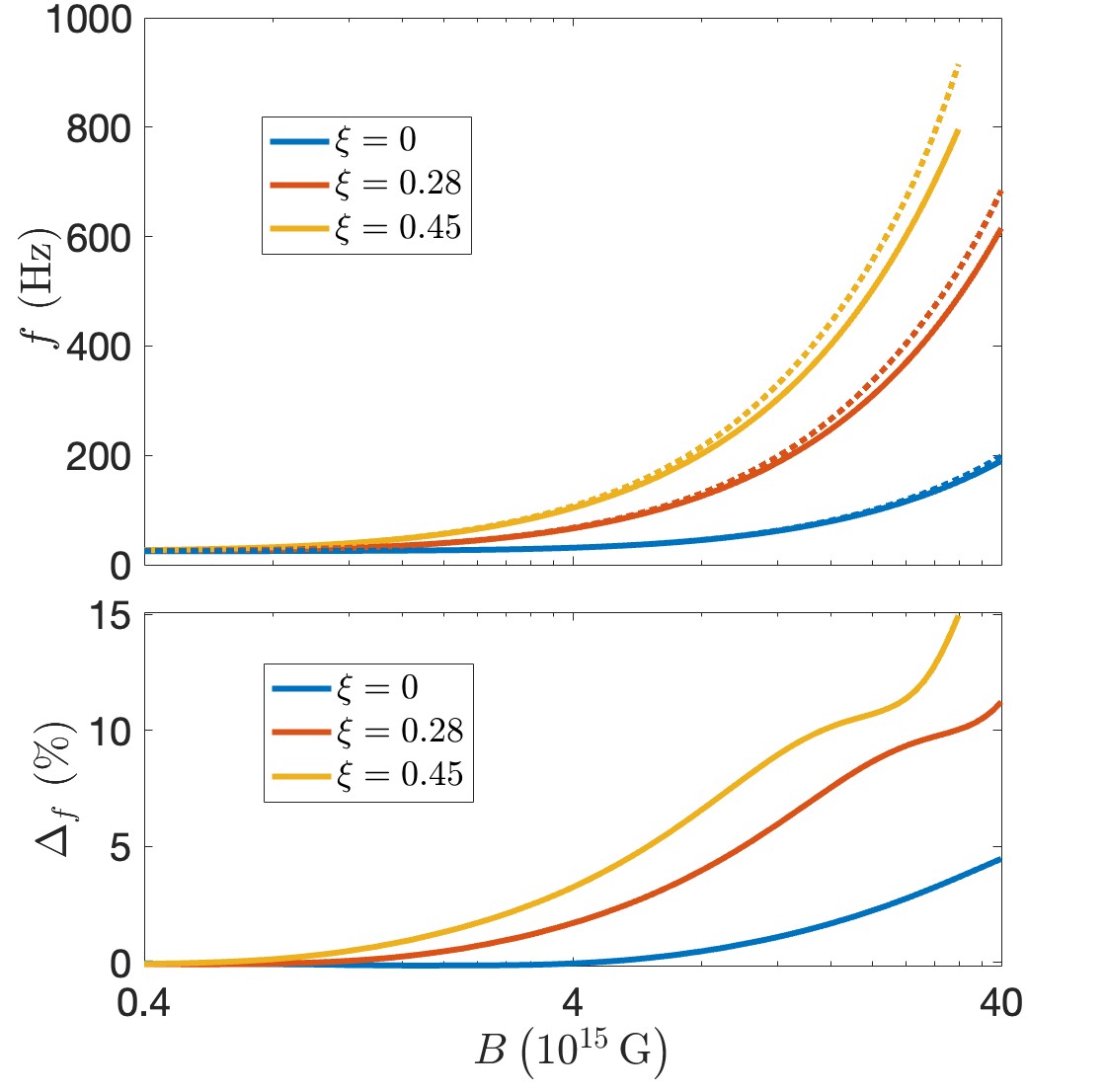}
    \caption{Upper panel: Mode frequency $f$ as a function of $B$ for $\xi = 0$, 0.28 and 0.45. The solid lines correspond to $f^{\circ}$, while the dotted lines correspond to $f'$. Lower panel: $\Delta_f$ as a function of $B$. The neutron star is assumed to have a normal fluid core and the magnetic field extends throughout the star.}
    \label{fig:combine_f_normal}
\end{figure}

In the upper panel of Fig.~\ref{fig:combine_f_normal}, we plot $f'$ (dotted lines) and $f^{\circ}$ (solid lines)
against $B$ for neutron stars with a normal fluid core and magnetic field configurations defined by $\xi=0$, 0.28 and 0.45. One can observe that $f^{\circ}$ increases rapidly with $B$ and $\xi$ when the field strength is larger 
than about $10^{15}$ G. This phenomenon has been discussed in \cite{Sotani2007,de_Souza_2019}. 
A semi-analytical understanding is provided in Appendix~\ref{sec: proof}.  
Including the effect of magnetic field on the EOS (dotted lines), the mode frequency further increases 
for fixed values of $B$ and $\xi$. 
The difference between $f'$ and $f^{\circ}$ also increases with the complexity (i.e., the value of $\xi$)
of the field configuration. The difference $\Delta_f$ is plotted in the lower panel of Fig.~\ref{fig:combine_f_normal}. For $B=4\times 10^{15}$ G, $\Delta_f$ can reach only up to a few percent
even for the case $\xi=0.45$, which is already near the upper limit of the physical range II defined in 
Eq.~(\ref{eq:Bfield_parameters}). However, $\Delta_f$ can rise up to about 10\% when the field strength 
$B \approx 2\times 10^{16}$ G for $\xi=0.28$ and 0.45. For the case of purely dipole field ($\xi=0$), $\Delta_f$
can reach only up to about 5\% even when $B=4\times 10^{16}$ G. 
As discussed in Sec.~\ref{sec:STRUCTURES OF NEUTRON STARS}, for a fixed value of $\xi$, including the effect of magnetic field on the EOS would effectively increase the complexity of the field configuration, and hence leading
to the increase of $\Delta_f$ as $B$ increases.  


\subsection{Superconducting cores}
\begin{figure}
	\includegraphics[width=\columnwidth]{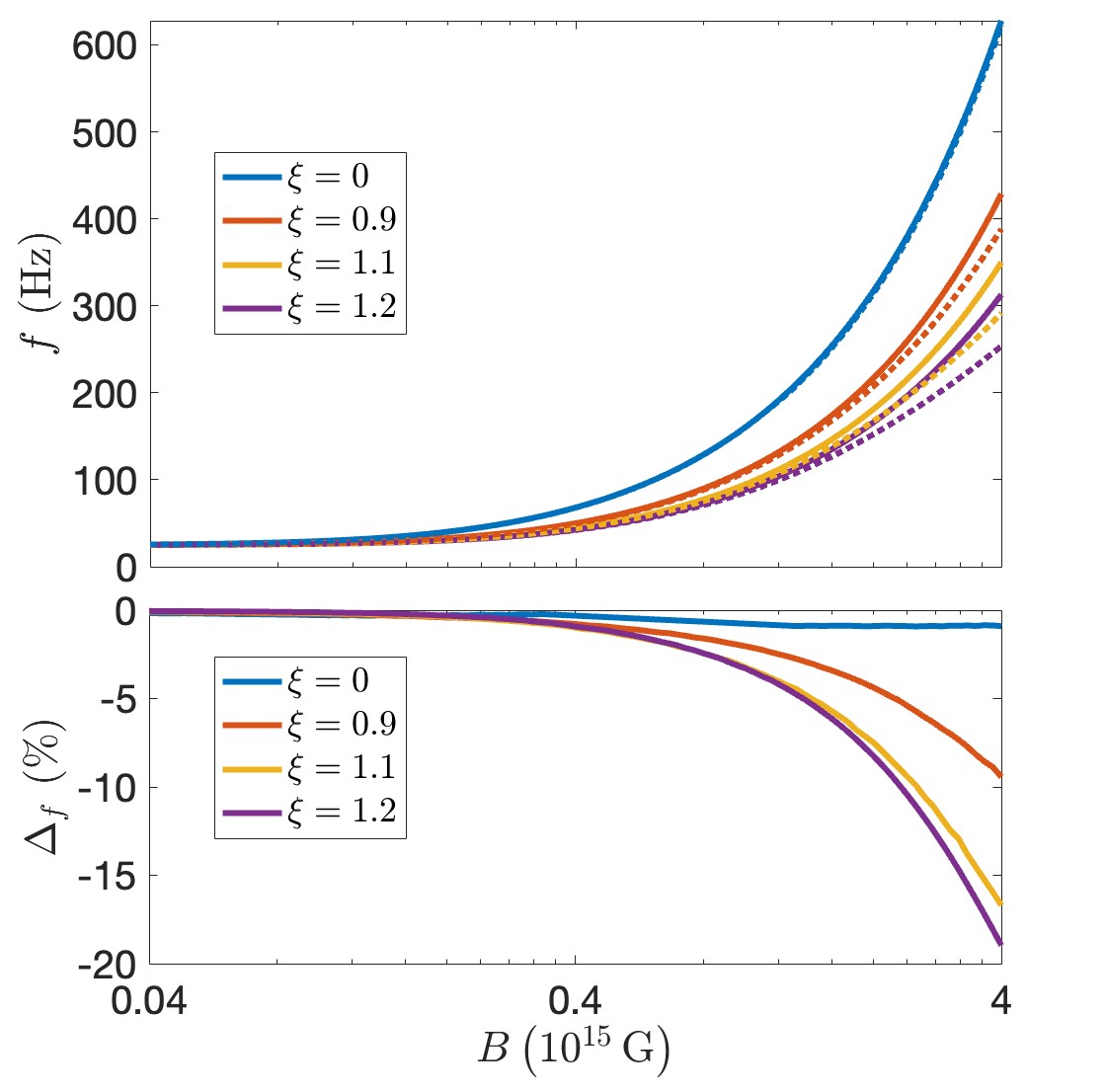}
    \caption{Upper panel: $f$ as a function of $B$ for $\xi = 0$, 0.9, 1.1 and 1.2. The solid and dotted lines correspond to $f^{\circ}$ and $f'$, respectively. Lower panel: $\Delta_f$ as a function of $B$. The neutron star is assumed to have a superconducting core and the magnetic field is confined in the crust.}
    \label{fig:combine_f_superfluid}
\end{figure}

For neutron stars with a superconducting core, the results differ from the above results qualitatively.
Similar to Fig.~\ref{fig:combine_f_normal}, we plot $f^{\circ}$ and $f'$ in the upper panel of 
Fig.~\ref{fig:combine_f_superfluid} to illustrate how the mode frequencies change with $B$ and $\xi$. 
In this case, it is seen that $f^{\circ}$ increases with $B$ but decreases with $\xi$. 
Hence, in contrast to neutron stars with a normal fluid core, increasing the complexity of the field configuration (i.e., the value of $\xi$) would decrease the mode frequency. A semi-analytical understanding is provided in Appendix~\ref{sec: proof}. 

Again, as discussed in Sec.~\ref{sec:STRUCTURES OF NEUTRON STARS}, including the effect of magnetic field on the EOS effectively increases the complexity of the field configuration for a fixed value of $\xi$, since the 
upper limit $\xi_{\gamma}$ of the physical range defined in Eq.~(\ref{eq:Bfield_xi_gamma}) decreases with 
increasing $B$. This is why the frequency $f'$ is smaller than $f^{\circ}$ for given values of $B$ and $\xi$. 
The difference $\Delta_f$ between the two frequencies increases with $B$ and $\xi$ as shown in the lower panel 
of Fig.~\ref{fig:combine_f_superfluid}. 
In particular, the magnitude of $\Delta_f$ can reach up to about 20\% when $B = 4\times 10^{15}$ G for the case $\xi=1.2$, which is close to the upper limit of $\xi_\gamma$ as shown in Fig.~\ref{fig:combine_structure_superfluid}. 

\subsection{Magnetic field complexities}
\begin{figure}
	\includegraphics[width=\columnwidth]{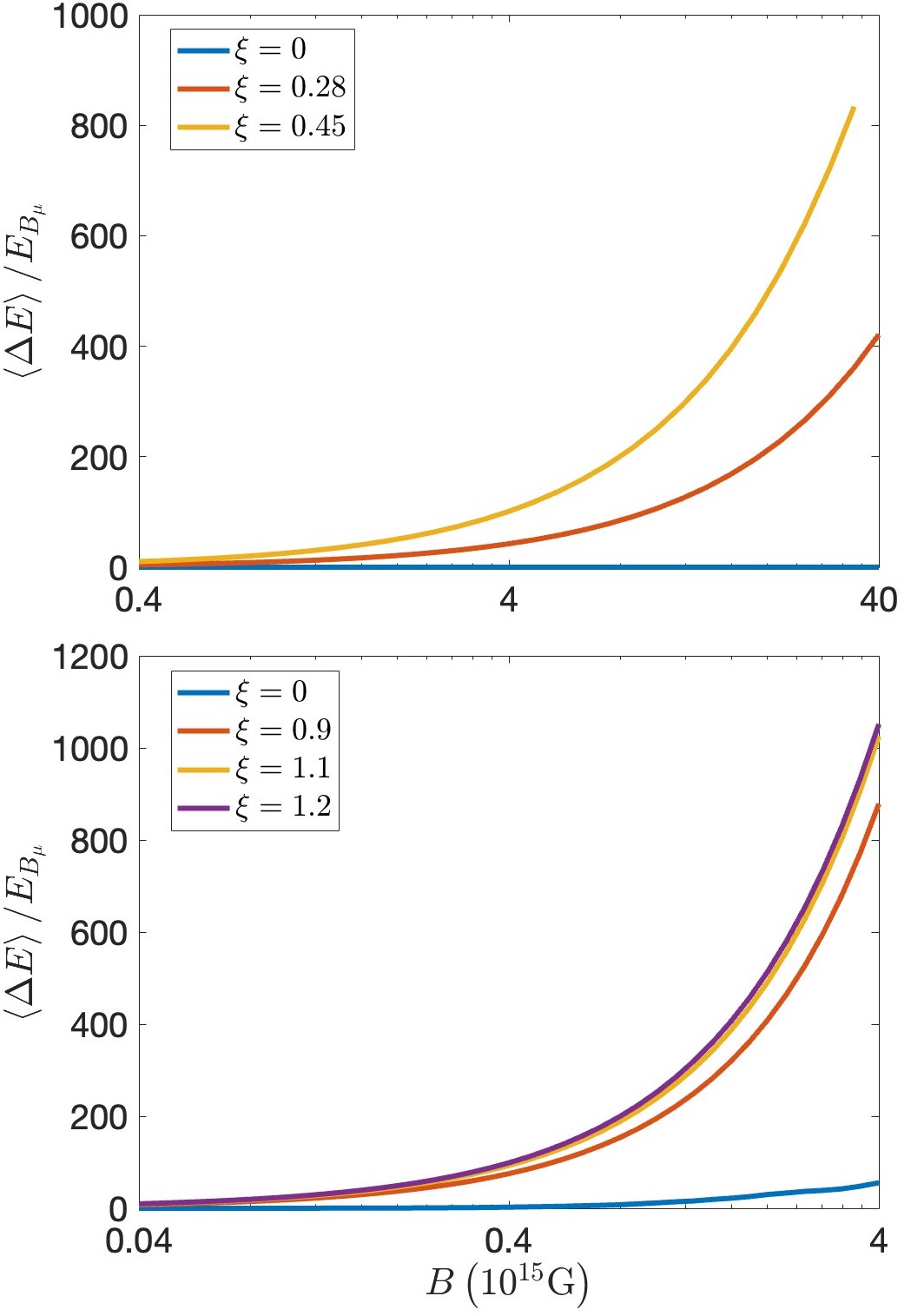}
\caption{$\langle \Delta E \rangle$ as a function of $B$ for $1.4 M_\odot$ neutron stars with a normal fluid (upper panel) and superconducting core (lower panel). Here we normalize $\langle \Delta E \rangle$ by $E_{B_\mu}$ for convenience, where $E_{B_\mu}$ is the magnetic energy density for the magnetic field strength $B_\mu = 4 \times 10^{15}$ G.}
    \label{fig:complexity_paper}
\end{figure}

In the above discussion, we characterize the complexity of a magnetic field configuration by the parameter $\xi$, which is compared to a purely dipole configuration ($\xi=0$) constructed with a TOV background without the effect of Landau-Rabi quantization (i.e., an unmagnetized background). 
We see that the value of $\xi$, and hence the complexity of the field configuration, plays an important role in the oscillation mode frequencies. 
However, $\xi$ is not intuitive and it may be better to compare a physical quantity associated with the magnetic energy. In order to better capture the 
difference between a generic field configuration and a purely dipole field on an unmagnetized background, we define the quantity 
\begin{equation}
\langle \Delta E \rangle = 
\sqrt{\left (\Delta E_r\right )^2 +\left (\Delta E_\theta\right )^2 + \left (\Delta E_\phi\right )^2} ,
\end{equation} 
where $\Delta E_i \left (i=r,\theta,\phi\right )$ stands for the difference of the magnitude
of $E_i$ between the two field configurations at the same value of $B$, in which $E_i$ is the mean magnetic energy density of the $i$-component of the magnetic field stored in the crust. 
In Fig.~\ref{fig:complexity_paper}, we plot $\langle \Delta E \rangle /E_{B_\mu}$ against $B$, 
where $E_{B_\mu}$ is the energy density for the 
field strength of $B_\mu = 4\times 10^{15}$ G (which is introduced in 
Eq.~(\ref{eq:sotani_fitting})) in the crust.
For a fixed value of $B$, Fig.~\ref{fig:complexity_paper} shows that $\langle \Delta E \rangle$ increases with $\xi$ for both the normal fluid (upper panel) and superconducting cores (lower panel). Hence, $\xi$ can serve as a proxy for the physical quantity $\langle \Delta E \rangle$ which quantifies the complexity of the field configuration relative to a purely dipole field on an unmagnetized background.

\section{Effects of stellar mass}
\label{sec:Another model for comparison}

So far we have focused on a $1.4 M_\odot$ neutron star. Let us now present the results for a more massive $2.0 M_\odot$ model with $\delta r^{\circ} = 0.4615$ km and $R^{\circ} = 12.2298$ km in the non-magnetized limit. We shall see how $\Delta_{\delta r}$ and $\Delta_f$ change for neutron stars with different masses. It is worth noticing that the mode frequency $f^\circ$ for a $2.0M_\odot$ neutron star model is not the same as that of the $1.4M_\odot$ model. For comparison, we also present the corresponding values of $\xi_a,\xi_b,\xi_c$ and $\xi_{\gamma}$ for $1.4M_\odot$ and $2.0M_\odot$ neutron star models in the weak field limit ($B\ll 10^{15}$ G) in Table~\ref{tab:compare_mass_xi}.

\subsection{Normal fluid cores}
\begin{figure}
	\includegraphics[width=\columnwidth]{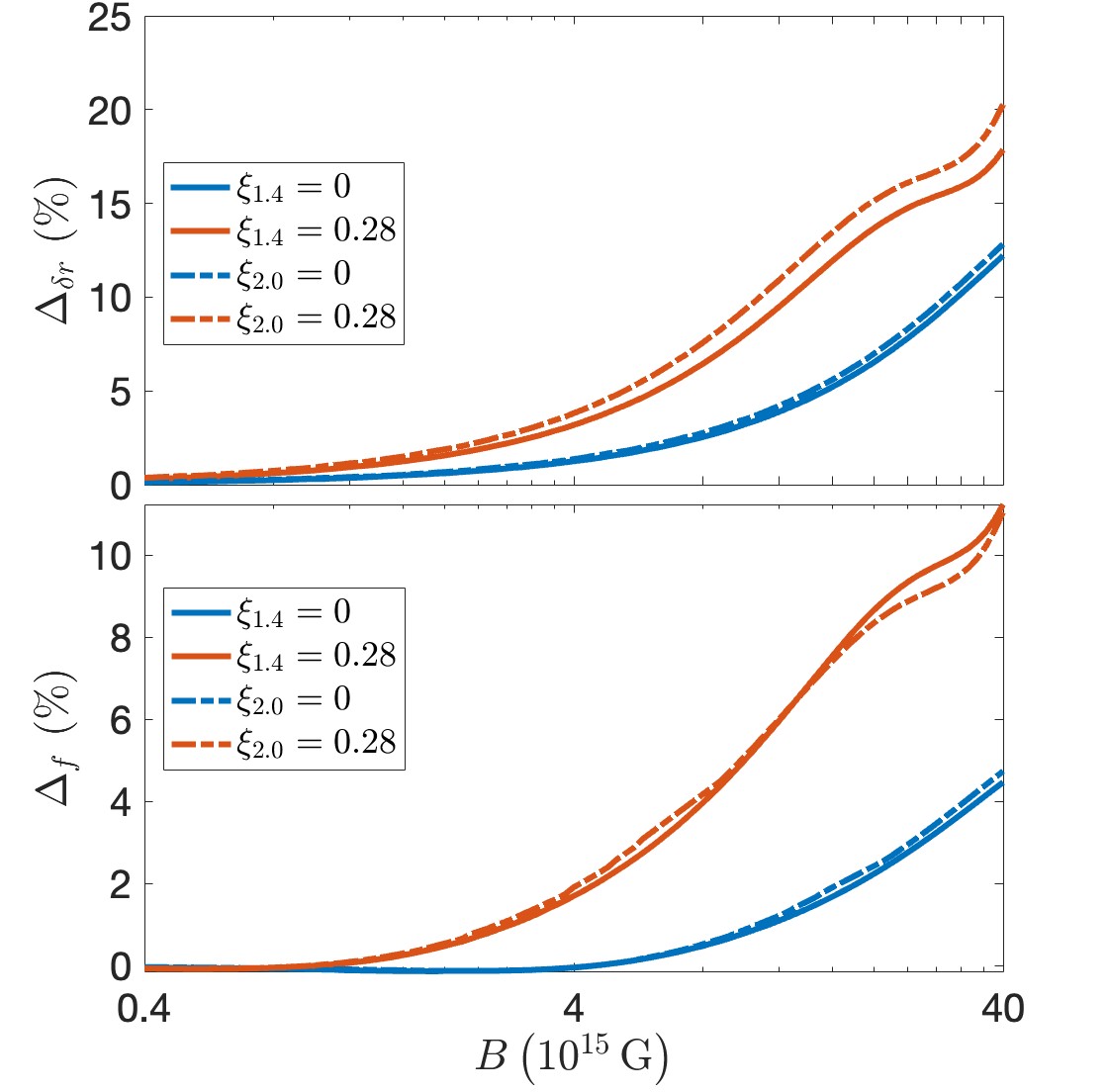}
    \caption{Upper panel: $\Delta_{\delta r}$ as a function of $B$ for $\xi = 0$ (blue lines) and 0.28
    (red lines). Lower panel: $\Delta_f$ as a function of $B$. The solid and dotted-dashed lines
    correspond to neutron star models with $1.4 M_\odot$ and $2.0 M_\odot$, respectively. The neutron 
    stars are assumed to have a normal fluid core, and the magnetic field extends throughout the stars. 
    }
    \label{fig:compare_mass}
\end{figure}
\begin{table}
	\centering
	\caption{The values of $\xi_a,\xi_b,\xi_c$ and $\xi_\gamma$ for $1.4M_\odot$ and $2.0M_\odot$ neutron star models in the weak field limit ($B \ll 10^{15}$ G).}
	\label{tab:compare_mass_xi}
	\begin{tabular}{lcccr} 
		\hline
		  $M/M_\odot$ & $\xi_a$ & $\xi_b$ & $\xi_c$&$\xi_\gamma$\\
		\hline
		1.4 & 0.298 & 0.354 & 0.462&1.4\\
		2.0 & 0.255 & 0.265 & 0.365&2.1\\
		
		\hline
	\end{tabular}
\end{table}

We first compare how the differences in the thickness of the crust $\Delta_{\delta r}$ defined in 
Eq.~(\ref{eq:Delta_dr}), relative to the nonmagnetic limit, depend on the stellar mass in the upper panel of 
Fig.~\ref{fig:compare_mass}. In the figure, the blue lines correspond to the purely dipole field configuration
($\xi=0$) for star models with $1.4 M_\odot$ (solid line) and $2.0 M_\odot$ (dotted-dashed line). 
It is seen that the two lines agree quite well with each other even up to $B=4 \times 10^{16}$ G, and hence 
$\Delta_{\delta r}$ is insensitive to the stellar mass for $\xi=0$. 
For a more complex field configuration defined by $\xi=0.28$ (red lines), the results are slightly more sensitive to the stellar mass. Similarly, we plot the corresponding differences in the mode frequency $\Delta_f$ in the lower panel of Fig.~\ref{fig:compare_mass} and notice that $\Delta_f$ is somewhat less sensitive to the stellar mass for $\xi=0.28$ comparing to $\Delta_{\delta r}$.

\subsection{Superconducting cores}

\begin{figure}
	\includegraphics[width=\columnwidth]{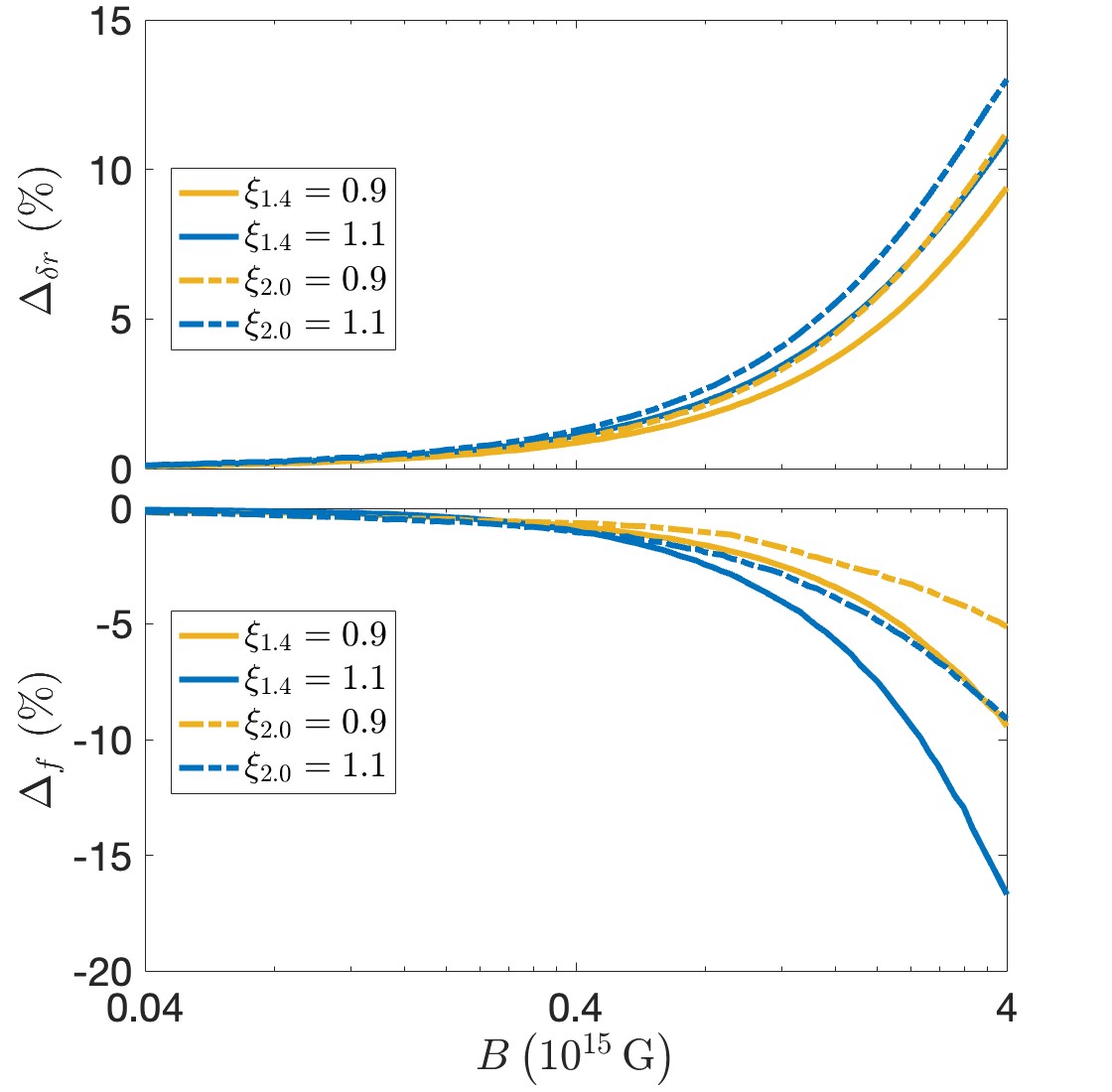}
    \caption{Upper panel: $\Delta_{\delta r}$ as a function of $B$ for $\xi = 0.9$ (yellow lines)
    and 1.1 (blue lines). Lower panel: $\Delta_f$ as a function of $B$. 
    The solid and dotted-dashed lines correspond to neutron star models with $1.4 M_\odot$ and
    $2.0 M_\odot$, respectively. The neutron stars are assumed to have a superconducting core, and the
    magnetic field is confined in the crust. }
    \label{fig:compare_mass_superfluid}
\end{figure}

For neutron stars with a superconducting core, the effects of stellar mass are different from those discussed above. 
Similar to Fig.~\ref{fig:compare_mass}, we consider how $\Delta_{\delta r}$ depends on stellar mass in the upper panel of Fig.~\ref{fig:compare_mass_superfluid}.
The yellow (blue) lines indicate the results for stellar models constructed by the field 
configuration $\xi=0.9$ ($\xi = 1.1$). The $1.4M_\odot$ and $2.0 M_\odot$ models are 
represented by the solid and dotted-dashed lines, respectively.
Unlike the situation for a normal fluid core, it is seen that $\Delta_{\delta r}$ is now more sensitive 
to the stellar mass as the blue solid and dotted-dashed lines, and similarly for the yellow lines,
already deviate from each other significantly when $B=4\times 10^{15} \textrm{G}$. 
Nevertheless, $\Delta_{\delta r}$ in general increases with the stellar mass for both normal fluid
and superconducting cores. 
In the lower panel of Fig.~\ref{fig:compare_mass_superfluid}, we plot the corresponding $\Delta_f$ for neutron stars with a superconducting core. 
In contrast to the case for a normal fluid core, where $\Delta_f$ is weakly dependent on the stellar 
mass, it is seen that $\Delta_f$ now depends more sensitively on the stellar mass. 
In particular, the magnitude of $\Delta_f$ increases with decreasing stellar mass for a given 
magnetic field strength.

\section{Discussion}
\label{sec:Discuss}

The intense magnetic field of a magnetar can affect the structure and EOS of the crust of 
the star. In this paper, we have studied the effects of Landau-Rabi quantization of electron motion on the crustal torsional oscillations of magnetars in general relativity under the Cowling approximation. For the EOS model, we employ the study of \citep{Mutafchieva_2019} which incorporates the effects of Landau-Rabi quantization by treating the inner and outer crusts consistently based on the nuclear-energy density functional theory. 
We also consider mixed poloidal-toroidal magnetic field configurations for our stellar models. 
In our study, the magnetic field configuration first contributes to the calculation of the torsional oscillation modes through its effect on the magnetic-field-dependent EOS due to the Landau-Rabi quantization, and hence the resulting magnetized stellar model as discussed in Sec.~\ref{sec:SELF-CONSISTENT STELLAR MODELS}. The magnetic field then contributes at the level of the perturbation equations that determine the oscillation modes.   
We assume spherically symmetric stellar models and ignore the crust deformation \citep{Franzon_2017} and the modification of the energy-momentum tensor due to the strong magnetic field \citep{Chatterjee_2021}.   
These effects are expected to be non-negligible only when the strength of magnetic field is of the order of $10^{18}$ G, which is far beyond our range of study here and the typical strength of magnetic field of magnetars.

The magnetic field configurations are characterized by the strength of the magnetic field
$B$ at the pole of the stellar surface and the parameter $\xi$, which represents the ratio 
between the toroidal and poloidal components of the magnetic field. Our numerical results show that the crust thickness $\delta r$ increases with both $B$ and $\xi$ when the effect of magnetic field is considered in the EOS. For a more complex magnetic field configuration with a nonzero $\xi$, $\delta r$ can increase by more than 20\% when $B \sim 10^{16}$ G for neutron stars with a normal fluid core. On the other hand, for neutron stars with a superconducting core, $\delta r$ can already increase by 10\% when $B \sim 10^{15}$ G. 

\begin{figure}
	\includegraphics[width=\columnwidth]{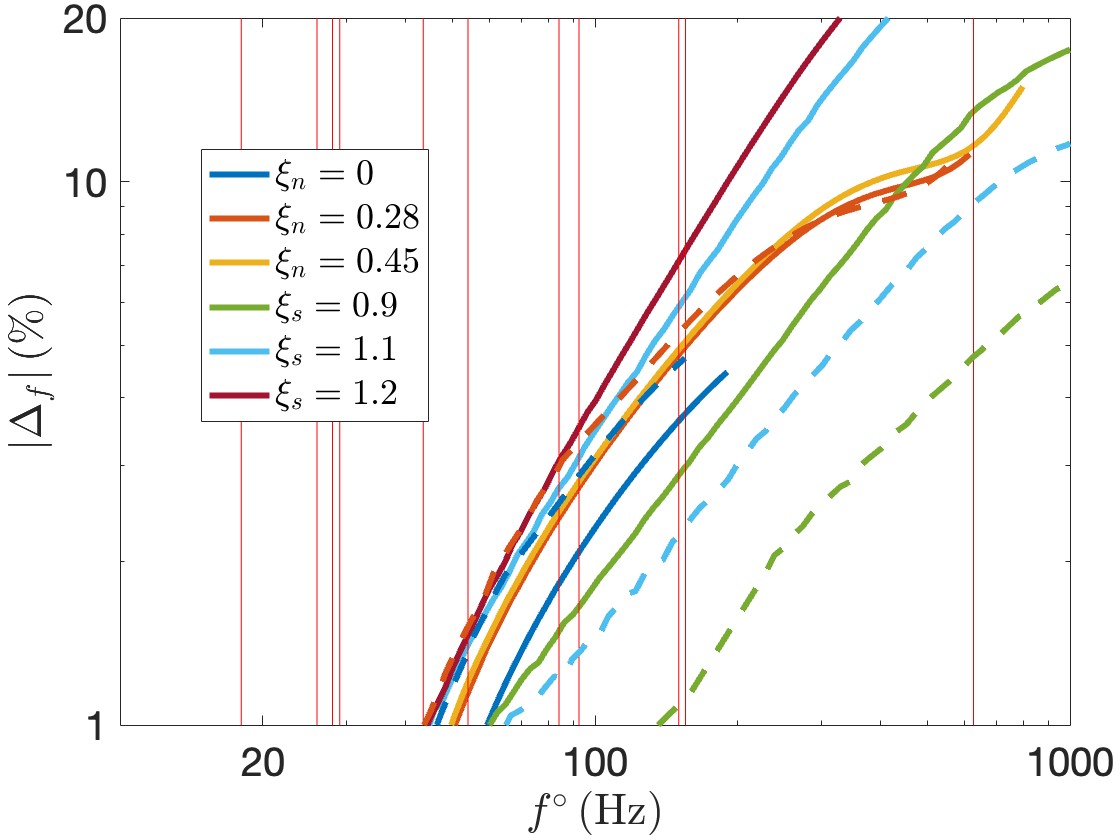}
    \caption{$\left |\Delta_{f}\right |$ as a function of $f^\circ $ for different models. 
    The subscripts $n$ and $s$ in the parameter $\xi$ indicate that the neutron star models consists of a 
    normal fluid and superconducting core, respectively. The solid lines and dashed lines represent the results for 1.4 $M_\odot$ and 2.0 $M_\odot$ neutron star models, respectively. The vertical lines denote some of the observed QPO frequencies in magnetar giant flares.  }
    \label{fig:observe}
\end{figure}

We focus on the fundamental $\ell=2$ torsional oscillation modes and study the changes in the mode frequencies when the effect of the magnetic field on the EOS is incorporated. It is found that the magnitude of the fractional change in mode frequency $\left |\Delta_f\right |$ increases with $B$ and $\xi$. For the models that we have considered, $|\Delta_f|$ can exceed 10\% when $B\sim 10^{16}$ G for neutron stars with a normal fluid core. For neutron stars with a superconducting core, however, $|\Delta_f |$ can even approach 20\% at a smaller field strength $B\sim 10^{15}$ G, which lies at the top end of the observed range of magnetic field for magnetars. It is also worth to note that $\Delta_f$ is positive (negative) for neutron stars with a normal fluid (superconducting) core, and hence the effect of Landau-Rabi quantization would increase (decrease) the mode frequency correspondingly. We have also studied the effect of stellar mass and found that $\Delta_f$ is not sensitive to the stellar mass for models with a normal fluid core.  However, for a superconducting core, $\Delta_f$ depends more sensitively on the stellar mass and increases with decreasing mass, especially for field strength larger than $B\sim 10^{15}$ G.


Let us now put together the results of all the stellar models studied in this work and discuss 
the implication from an observational perspective. While the mechanism for the magnetar giant flares and the 
nature of the observed QPOs in the late-time tail phase are still not well understood, it is believed that the magneto-elastic oscillations of the star must play a role.   
Although the role of torsional oscillation modes of magnetars   
in explaining (at least) some of the observed QPOs is far from clear, let us 
adopt this simple scenario as an example to illustrate the implication of our numerical results. In particular, we want to clarify the situation under which we need 
to worry about the effect of Landau-Rabi quantization and the level of deviations one could have if 
this effect is not considered in the theoretical calculation of the oscillation modes, when comparing
to the observations. 

In Fig.~\ref{fig:observe}, we plot $\left |\Delta_{f}\right |$ as a function of the mode 
frequencies $f^\circ $ without considering the effect of magnetic field on the EOS for different models
studied in this work, where the subscripts $n$ and $s$ in the parameter $\xi$ indicate that the neutron 
star models consist of a normal fluid and superconducting core, respectively. The solid and dashed lines
with the same color represent $1.4 M_\odot$ and $2.0 M_\odot$ models for the same value of $\xi$, respectively. In addition, the vertical lines denote the frequencies of some of the observed QPO as 
mentioned in Sec.~\ref{sec:intro}.
Figure~\ref{fig:observe} shows that the effect of Landau-Rabi quantization is not important when one
considers the observed QPOs with smaller frequencies ($\lesssim 80$ Hz), but becomes relevant  
for higher frequencies modes. It can be seen that $\left |\Delta_f\right |$ is always less than 1\% for 
$f^\circ$ being around 40 Hz. However, $\left |\Delta_f \right |$ exceeds 3\% for $f$ being around 90 Hz and approaching 10\% for $f$ being around 150 Hz.

\begin{figure}
	\includegraphics[width=\columnwidth]{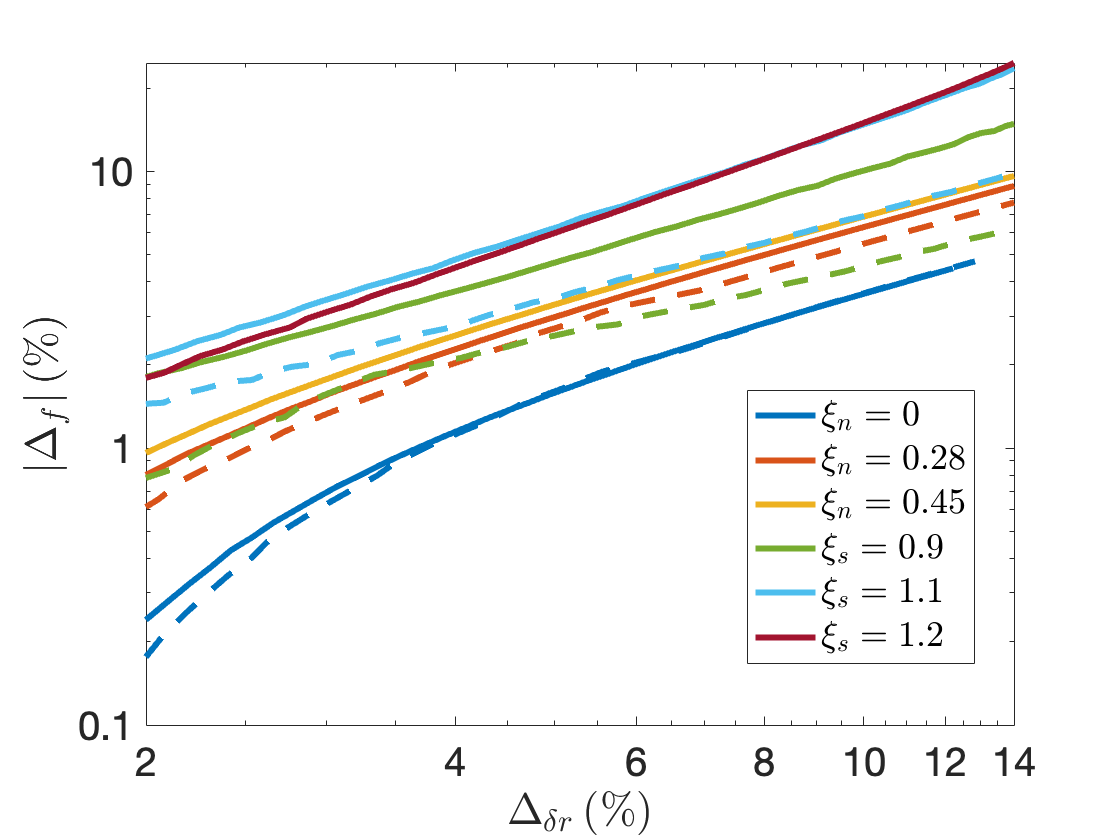}
    \caption{$\left |\Delta_{f}\right |$ as a function of $\Delta_{\delta r} $ for different models, similar to Fig.~\ref{fig:observe}.    }
    \label{fig:deltar_deltaf}
\end{figure}

In addition, one can see that different models show similar trends in Fig.~\ref{fig:observe}, which can be 
fitted approximately by 
\begin{equation}
\label{eq:fitting1}
    \left | \Delta_f \right |=K\left (  f^\circ \right )^{1.12\pm0.12},
\end{equation}
where $K$ is a constant depending on the background star model. This fit applies to the range for 
$\left | \Delta_f \right |$ being between 1\% and 10\%. 
On the other hand, we plot $\left | \Delta_f \right |$ against $\Delta_{\delta_r}$ in Fig. \ref{fig:deltar_deltaf}. As one may expected, there should be some correlation between the two quantities. 
Different models also show similar trends in Fig.~\ref{fig:deltar_deltaf}, which can be fitted 
for $\Delta_{\delta r}$ in the range between 2\% and 10\% approximately by
\begin{equation}
\label{eq:fitting2}
    \left | \Delta_f \right |=K'\left (  \Delta_{\delta r} \right )^{1.11\pm0.04},
\end{equation}
where $K'$ is a constant depending on the background star model. 



In conclusion, depending on the neutron star models and magnetic field configurations, we have shown that the effect of Landau-Rabi quantization of electron motion in the crust can change the frequencies of the fundamental
torsional oscillation modes by up to 10\% to 20\% for magnetic field strength $B \sim 10^{15} - 10^{16}$ G,
which is within the measured surface field strength of magnetars.
However, it should be noted that all the results presented here 
only hold in cases where the important coupling to the core and a more realistic treatment of the boundary conditions are not considered.
Our work provides a benchmark on the parameter space of stellar models and observation precision within which
one needs to consider the Landau-Rabi quantization. 
In the future, we plan to extend our study to other 
oscillation modes such as the interfacial and shear modes, which also depend sensitively on the properties of 
the solid crust and may be excited in magnetar starquakes. 


\section*{Acknowledgements}

The work of NC was financially supported by the Fonds de la Recherche Scientifique (Belgium) under Grant No. PDR T.004320. LML is supported by a grant from the Research Grants Council of the Hong Kong SAR, China [Project No: CUHK 14304322].

\appendix

\section{CODE TESTS WITH MIXED POLOIDAL-TOROIDAL MAGNETIC FIELDS}
\label{sec:Code test figure}

\begin{figure}
	   \includegraphics[width=\columnwidth]{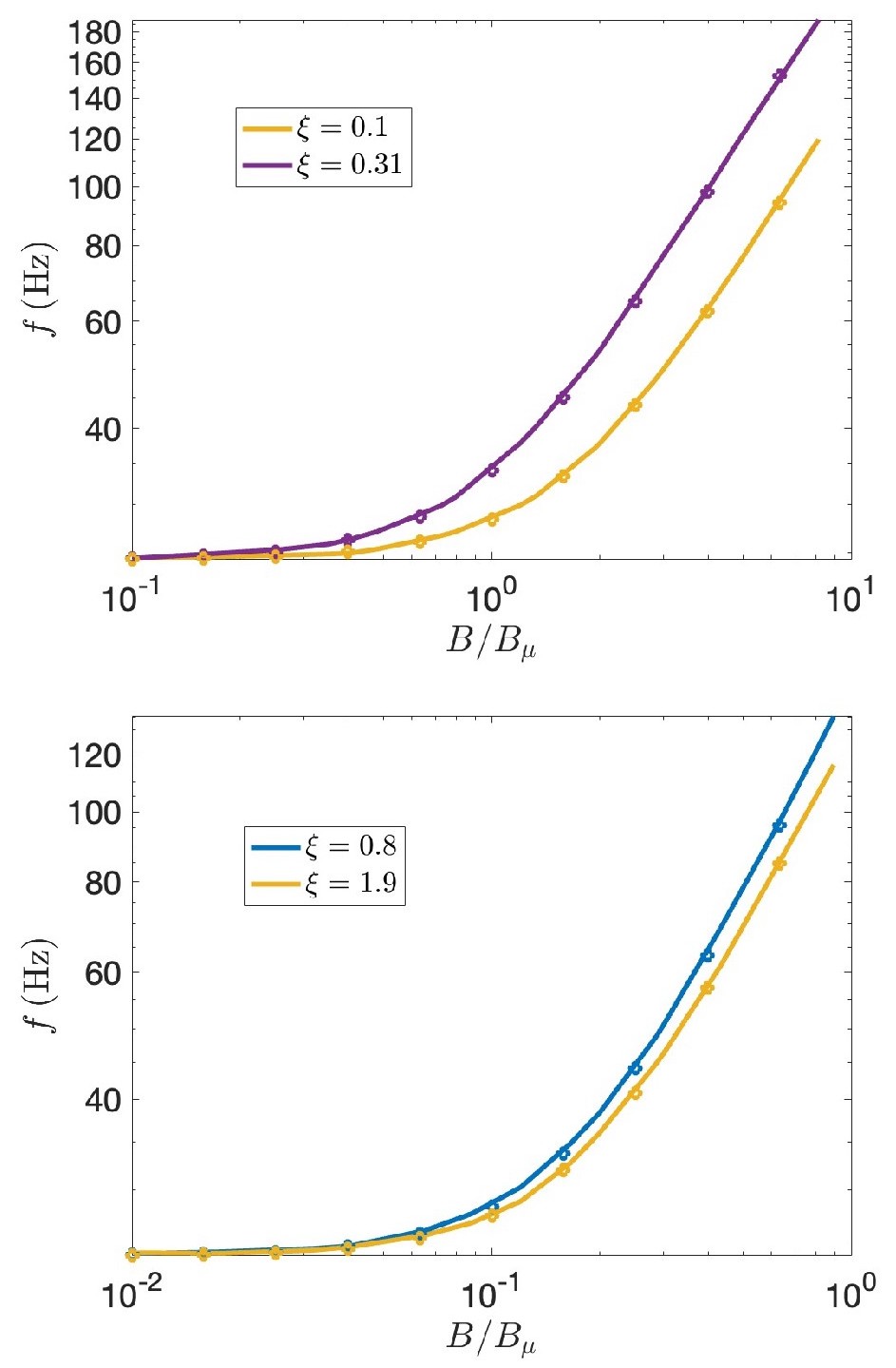}
    \caption{Comparison between our numerical results (data points) and the results (solid lines) 
    presented in \citep{Sotani_2008_phi} for a $1.4 M_\odot$ neutron star model with different 
    magnetic field configurations (see text for more details). The upper (lower) panel plots the   
    fundamental torsional oscillation mode frequency $f$ against the normalized field strength 
    $B/B_\mu$ for the case of a normal fluid (superconducting) core.}
    \label{fig:combined_code_test}
\end{figure}

In this appendix, we verify our code by comparing our numerical results for torsional oscillation 
modes with mixed poloidal-toroidal magnetic field configurations with the data published in 
\cite{Sotani_2008_phi}. We consider their 1.4 $M_\odot$ stellar model described by the following
EOS: 
\begin{equation}
    P=0.00936n_0m_b\left (\frac{\bar{n}}{n_0}\right )^{2.46},
\end{equation}
\begin{equation}
    \rho=\bar{n}m_b+\frac{P}{1.46},
\end{equation}
\begin{equation}
    \mu=\rho v_s^2,
\end{equation}
where $n_0=0.1\,\textrm{fm}^{-3}$, $m_b=1.66\times 10^{-24}\textrm{g}$ and $v_s = \sqrt{\mu/\left (\rho + P\right )}$ is the velocity of the shear wave as measured by a local observer which is assumed here to be fixed at its typical value of $10^8\,\textrm{cm\,s}^{-1}$ \cite{SchumakerThorne1983}.
In the upper panel of Fig. \ref{fig:combined_code_test}, we plot the fundamental torsional oscillation
mode frequencies $f$ against the normalized magnetic field strength $B/B_\mu (B_\mu = 4\times 10^{15}
{\ \rm G})$ for this background star with a normal fluid core and magnetic field configurations defined by $\xi=0.1$ and 0.31. The solid lines are the results extracted from the figures in \cite{Sotani_2008_phi} and the dots are our numerical results.  
Similarly, the lower panel of Fig. \ref{fig:combined_code_test} shows the comparison for neutron stars with a superconducting core and magnetic field configurations defined by $\xi = 0.8$ and 1.9.
We see that our results match those presented in \cite{Sotani_2008_phi} very well.

\section{SEMI ANALYTICAL UNDERSTANDING OF TORSIONAL OSCILLATION MODES WITH MIXED POLOIDAL-TOROIDAL MAGNETIC FIELDS}
\label{sec: proof}

To understand the behaviours of the torsional oscillation mode frequency $f$ with mixed poloidal-toroidal magnetic fields, we can apply approximations on Eq.~(\ref{eq:pulsation equation}). In this appendix, 
we do not consider the effects of magnetic field on the EOS.

\subsection{Normal fluid core}

Let us first consider neutron stars with a normal fluid core. To simplify the problem, we only focus on the Newtonian limit. As the torsional oscillations are confined in the solid crust, we only need to analyze the 
crustal region. In addition, thin crust will be assumed in the following. 
For a thin crust, we have $d^2Y/d r^2\approx dY /d r\approx 0$ in our analysis due to the boundary conditions 
at the base of the crust and the stellar surface, which require $Y_2= dY /d r=0$. 

As we are only interested in how the mode angular frequency $\sigma$ scales with the magnetic field strength
$B$ and the parameter $\xi$ qualitatively, we can perform an order-of-magnitude approximation for 
Eq.~(\ref{eq:Y_2_equation}) and obtain the 
following result:
\begin{equation}
\begin{split}
\sigma^2 \approx&\frac{(\ell-1)(\ell+2)}{ R^2} \left \langle v_s^2\right \rangle\\
&+\frac{2+5\lambda_1 }{2 \pi \left \langle \rho \right \rangle R^4}\left [ \left \langle 4\pi f_0 \rho  R^2 a_1\right \rangle+\xi^2\left \langle a_1^2  \right \rangle \right ],
\end{split}
\end{equation}
where $\left \langle x \right \rangle$ is the average value of $x$ throughout the crust. 
In the absence of magnetic field (i.e., $a_1=0$), we obtain the non-magnetic eigenvalue
$\sigma^2 \approx {\bar\sigma}^2 \equiv (\ell-1)(\ell+2) \left \langle v_s^2\right \rangle/ R^2$, which agrees with the result obtained in 
\cite{Samuelsson:2007} except that a metric function of order unity is ignored in our analysis. We thus have
\begin{equation}
\begin{split}
\frac{\sigma^2}{\bar{\sigma}^2} \approx  1
+\frac{2+5\lambda_1}{2\pi \bar{\sigma}^2 \left \langle \rho \right \rangle  R^4}\left [ \left \langle 4\pi f_0 \rho  R^2 a_1\right \rangle+\xi^2\left \langle a_1^2  \right \rangle \right ].
\end{split}
\end{equation}

On the other hand, the source term $f_0\left (\rho + P\right )R^2$ in Eq.~(\ref{eq:solve b-field}) 
scales like $B$, as that is the current distribution producing the magnetic field. 
As $a_1 \propto B$ at the surface, we can approximate $\left \langle a_1^2 \right \rangle \propto B^2$ and
obtain the final relation
\begin{equation}
\begin{split}
\frac{\sigma^2}{\bar{\sigma}^2}=1 + C_\ell B^2 + D_\ell B^2 \xi^2,
\end{split}
\end{equation}
where $C_\ell$ and $D_\ell$ are some positive constants depending on $\ell$ and the stellar models. Rewriting the terms give
\begin{equation}
\begin{split}
\frac{\sigma}{\bar{\sigma}}=\left [1 + C_\ell(1+E_\ell \xi^2)B^2 \right ]^{1/2},
\end{split}
\end{equation}
where $E_\ell=D_\ell/C_\ell$. This is exactly the results provided numerically in \cite{de_Souza_2019}. For purely dipole field ($\xi = 0$), it reduces to the fitting formula proposed in \cite{Sotani2007} 
(i.e., Eq.~(\ref{eq:sotani_fitting})).

\subsection{Superconducting core}
As stated in \cite{Sotani_2008_phi}, the $\theta$-component of magnetic field will be much stronger than the other components when the magnetic field is confined only in the crustal region. So, here we repeat the same 
analysis above but keep the terms with $\left (da_1/dr\right )^2$ instead of $a_1^2$ (see Eq.~(\ref{eq:B_component})). 
The final result can be written as
\begin{equation}
\begin{split}
 \frac{\sigma^2}{\bar{\sigma}^2} \approx 1+C_\ell B^2+ F_\ell \left \langle \left (\frac{da_1 }{d r}\right )^2\right \rangle,
\end{split}
\label{eq:sigma_super_App}
\end{equation}
where $F_\ell = \left ( \ell-1 \right )\left ( \ell+2 \right )\left ( \frac{\left |\lambda_1 \right |}{2\pi \bar{\sigma}^2\left \langle \rho \right \rangle R^4} \right )$. Applying similar approximations as above and only considering the cases with $\xi \sim 1$ for our numerical results, Eq.~(\ref{eq:solve b-field}) can 
be approximated by
\begin{equation}
\begin{split}
\frac{d^2a_1}{dr^2}+\frac{2M}{R^2}\frac{da_1}{dr}+\xi^2a_1 \approx \Gamma B,
\label{eq:xi_superfluid_approximation}
\end{split}
\end{equation}
where $\Gamma$ is some constant depending on the stellar model. We have also neglected the term $2a_1/r^2$ 
in Eq.~(\ref{eq:solve b-field}) in the analysis as here it is much smaller than the other terms.
In the following, $M/R^2$ will be taken as $0.01$ in our order-of-magnitude analysis. Solving this second-order ordinary differential equation, we obtain approximately
\begin{equation}
\begin{split}
a_1\left (r \right )\approx\chi_1 e^{-0.01r} \sin \left (\xi r \right ) + \chi_2 e^{-0.01r} \cos \left (\xi r \right ) + \frac{\Gamma B}{\xi^2},
\end{split}
\end{equation}
where $\chi_1$ and $\chi_2$ are some constants. By requiring $a_1\left (R-\delta r \right ) = \frac{d^2 a_1}{dr^2}\left (R-\delta r \right ) = 0 $ (see Sec.~\ref{sec:MIXED POLOIDAL-TOROIDAL MAGNETIC FIELDS} for the boundary conditions), both $\chi_1$ and $\chi_2$ scale like $B/\xi^2$. As a result, 
\begin{equation}
\begin{split}
\left \langle \frac{da_1}{dr}\right \rangle  \propto  \frac{B}{\xi}.
\end{split}
\end{equation}
Eq.~(\ref{eq:sigma_super_App}) can then be expressed as
\begin{equation}
\begin{split}
 \frac{\sigma^2}{\bar{\sigma}^2}= 1+C_\ell B^2+ G_\ell  \frac{ B^2}{ \xi^2},
\end{split}
\end{equation}
where $G_\ell$ is some constant depending on $\ell$ and the stellar model. We have tested this relation with
our numerical results to verify that the $\xi$-dependence obtained is valid and the error is less than 10\% for the range of magnetic field strength considered. Therefore, given a fixed value of $B$, increasing the value of $\xi$ will decrease $\sigma$.

For the cases with $\xi = 0$, the solution of Eq.~(\ref{eq:xi_superfluid_approximation}) is given 
approximately by
\begin{equation}
\begin{split}
a_1\left (r \right )\approx \chi_1 e^{-0.01r} + \chi_2  +100\Gamma B r.
\end{split}
\end{equation}
Again, having $a_1\left (R-\delta r \right ) = \frac{d^2 a_1}{dr^2}\left (R-\delta r \right ) = 0 $, we determine that $\chi_1=0$ and $\chi_2$ is just some constant. Therefore, we have $\left \langle \frac{da_1}{dr}\right \rangle  \propto  B$ and  
\begin{equation}
\begin{split}
 \frac{\sigma^2}{\bar{\sigma}^2}= 1+ J_\ell  B^2,
\end{split}
\end{equation}
where $J_\ell$ is some constant depending on $\ell$ and the stellar model. This relation has been tested with our numerical results and the error is around 1\% only. 


\bibliography{apssamp}

\end{document}